\documentclass[10pt,twoside,twocolumn]{IEEEtran}
\usepackage[latin9]{inputenc}
\usepackage[active]{srcltx}
\usepackage{color}
\usepackage{float}
\usepackage{amsmath}
\usepackage{amssymb}
\usepackage{graphicx}

\makeatletter

\floatstyle{ruled}
\newfloat{algorithm}{tbp}{loa}
\providecommand{\algorithmname}{Algorithm}
\floatname{algorithm}{\protect\algorithmname}

\usepackage{graphics}\usepackage{amsthm}\usepackage{amsopn}\usepackage{amstext}\usepackage{amsfonts}
\usepackage{bm}\usepackage{placeins}

\usepackage[noadjust]{cite}\newtheorem{theorem}{Theorem}
\newtheorem{proposition}[]{Proposition}
\newtheorem{lemma}[]{Lemma}
\newtheorem{definition}[]{Definition}
\newtheorem{remark}[]{Remark}

\newcounter{algo}

\newcommand{\wh}{\widehat}
\newcommand{\wt}{\widetilde}

\newcommand{\vect}[1]{\mathbf{#1}}

\ifCLASSINFOpdf
\else
\fi
\usepackage[footnotesize]{caption}

\makeatother

\begin{document}

\title{\vspace{-0.3cm}
 A New Decomposition Method for Multiuser DC-Programming and its Applications}

\author{Alberth Alvarado, Gesualdo Scutari, and Jong-Shi Pang \vspace{-0.9cm}
\thanks{Alvarado is with the ISE Dpt., Univ. of Illinois at Urbana-Champaign. Pang is with  the ISE Dpt.,
Univ. of Southern California Viterbi School of Eng., Los Angeles, USA. Scutari is with the Dpt.
of Electrical Eng., SUNY Buffalo. Emails: \texttt{alvarad3@illinois.edu},
\texttt{gesualdo@buffalo.edu}, \texttt{jongship@usc.edu}. 
\vspace{-0.8cm}
}}
\maketitle
\begin{abstract}
We propose a novel decomposition framework for the distributed optimization
of Difference Convex (DC)-type \emph{nonseparable} sum-utility functions
subject to \emph{coupling} convex constraints. 
 A major contribution of the paper is to develop for the first time
a class of (inexact) best-response-like algorithms with provable convergence,
where a suitably convexified version of the original DC program is
iteratively solved. The main feature of the proposed successive convex
approximation method is its decomposability structure across the users,
which leads naturally to \emph{distributed} algorithms in the primal
and/or dual domain. The proposed framework is applicable to a variety
of multiuser DC problems in different areas, ranging from signal processing,
to communications and networking. 

As a case study, in the second part of the paper we focus on two examples,
namely: i) a novel resource allocation problem in the emerging area
of cooperative physical layer security; ii) and the renowned sum-rate
maximization of MIMO Cognitive Radio networks. Our contribution in
this context is to devise a class of easy-to-implement distributed  algorithms
with \emph{provable convergence} to stationary solution of such problems.
Numerical results show that the proposed distributed schemes reach
performance close  to (and sometimes better than) that of centralized methods.

\vspace{-0.3cm}
 
\end{abstract}
\IEEEpeerreviewmaketitle

\section{Introduction }

\IEEEPARstart{T}{he} resource allocation problem in multiuser
systems generally consists of optimizing the (weighted) sum of the
users' objective functions, also termed a ``social function''.
In this paper we address the frequent and difficult case in which
the social function is nonconvex and there are (convex) shared constraints
coupling the strategies of all the users. Our attention is mainly
focused on objective functions of the DC-type, i.e., the difference of
two convex functions. It is worth mentioning that DC programs are
very common in signal processing, communications, and networking.
For instance, the following resource allocation problems belong to
the class of DC programs: power control problems in cellular systems
\cite{DC-BranchAndBound-1,DC-BranchAndBound-5,DC-Linearization-1};
MIMO relay optimization \cite{DC-Polynomial}; dynamic spectrum management
in DSL systems \cite{DC-BranchAndBound-2,TsiaflakisMoonenTSP08};
sum-rate maximization, proportional-fairness and max-min optimization
of SISO/MISO/MIMO ad-hoc networks \cite{Kim2011,SchmidtShiBerryHonigUtschick-SPMag,HongLuoTutorial12,DDPA-2}. 

In the effort of (optimally) solving DC programs, a great deal of
the aforementioned works involves global optimization techniques whose
solution methods are mainly based on combinatorial approaches (e.g.,
adaptations of branch and bound techniques \cite{DC-intro-1}); the
results are a variety of \emph{centralized }algorithms customized
to the specific DC structure under considerations \cite{DC-BranchAndBound-1,DC-BranchAndBound-2,DC-BranchAndBound-5,DC-Linearization-1,DC-Polynomial}.
However, centralized schemes are too demanding in most applications
(e.g., large-scale decentralized networks). This has motivated a number
of works whose effort has been finding efficiently high quality (generally
locally optimal) solutions of DC programs via easy-to-implement distributed
algorithms. Distributed\emph{ ad-hoc} schemes (with provable convergence)
for very specific DC formulations have been proposed in \cite{SchmidtShiBerryHonigUtschick-SPMag,Kim2011,DDPA-2,TsiaflakisMoonenTSP08},
mainly based on Successive Convex Approximation (SCA) techniques \cite{DDPA-2,SCA-3,DC-intro-3}.
In these works however the  formulations contain only \emph{private}
constraints (i.e., there is no coupling among the users' strategies). 

In this paper, we move a step forward and consider the more general\emph{
}multiuser DC program, whose feasible set includes also \emph{coupling}
convex constraints. To the best of our knowledge, the design of distributed
algorithms for this class of problems is an open issue. 
Indeed, the nonconvexity of the social function prevents the application
of standard primal/dual decomposition techniques for convex problems,
e.g., \cite{ConvexSumSeparable-1,ConvexSumSeparable-3}; the presence
of coupling constraints makes the distributed techniques developed
in \cite{SchmidtShiBerryHonigUtschick-SPMag,Kim2011,DDPA-2,TsiaflakisMoonenTSP08,SCA-3}
not directly usable; and standard SCA methods for DC programs \cite{DC-intro-3}
wherein the concave part of the objective function is linearized would
lead to centralized schemes (because the resulting convex function
is generally not separable in the users' variables). 

A first contribution of this paper is to develop a novel \emph{distributed}
decomposition method for solving such a class of multiuser DC problems.
Capitalizing on the SCA idea, the proposed novel technique solves
a sequence of strongly convex subproblems, whose objective function
is obtained by \emph{diagonalization plus off-diagonal linearization
of the convex part} of the original DC sum-utility and \emph{linearization
of the nonconvex part}. Some desirable features of the proposed approach
are: i) Convergence to a stationary solution of the original DC programming
is guaranteed also if the subproblems are solved in an inexact way;
ii) Each convex subproblem can be distributively solved by the users
capitalizing on standard primal or dual decomposition techniques;
and iii) It leads to alternative distributed algorithms that differ
from rate and robustness of convergence, scalability, local computation
versus global communication, and quantity of message passing. All
these features make the proposed technique and algorithms applicable
to a variety of networks scenarios and problems.

As a case study, in the second part of the paper we show how to customize
the developed framework to two specific problems, namely: 1) a novel
resource allocation problem in the emerging area of cooperative physical
layer security; and 2) the renowned sum-rate maximization of MIMO
Cognitive Radio (CR) networks. A brief description of our main contributions
in these two contexts is given next.

\subsubsection{Cooperative physical layer security}

Physical layer security has been considered as a promising technique
to prevent illegitimate receivers from eavesdropping on the confidential
message transmitted between intended network nodes; see, e.g., \cite{Jorswieck2010}
and references therein. Recently, cooperative transmissions using
trusted relays or friendly jammers to improve physical layer security
has attracted increasing attention \cite{CooperativeJamming-7,CooperativeJamming-8,CooperativeJamming-Petropulu2,GameTheoreticalModel-5,GameTheoreticalModel-6,GameTheoreticalModel-2,CooperativeJamming-Ma,li2013transmit}.
Of particular interest to this work is the Cooperative Jamming (CJ)
paradigm (see, e.g., \cite{CooperativeJamming-7,CooperativeJamming-8,CooperativeJamming-Petropulu2}):
friendly jammers create judicious interference by transmitting noise
(or codewords) so as to impair the eavesdropper's ability to decode
the confidential information, and thus, increase secure communication
rates between legitimate parties. The interference from the jammers
however might also reduce the useful rate of the legitimate links;
therefore the maximization of the users' secrecy rate calls for a
\emph{joint} optimization of the power allocations of the sources
\emph{and} the jammers. 

In this paper we address such a joint optimization problem. We consider a
network model composed of \emph{multiple} transmitter-receiver pairs,
\emph{multiple} friendly jammers, and a single eavesdropper. Note
that previous works studied simpler system models, composed of\textcolor{black}{{}
either }\textcolor{black}{\emph{one }}\textcolor{black}{source-destination
link (and possibly multiple jammers) or multiple sources but }\textcolor{black}{\emph{one}}\textcolor{black}{{}
jammer.} We formulate the system design as a game wherein the players$-$the
legitimate users$-$maximize their own secrecy rate by choosing \emph{jointly}
their transmit power and (the optimal fraction of the) power of the
friendly jammers. The resulting secrecy game faces two main challenges,
namely: i) the players' objective functions are nonconcave and nondifferentiable;
and ii) there are side (thus coupling) constraints. All this makes
the analysis of the proposed game a difficult task; for instance,
a Nash Equilibrium may not even exist. Capitalizing on recent results
on nonconvex games with side constraints \cite{nonconvexgame,Pang-Scutari-NNConvex_PI},
we introduce a novel relaxed equilibrium concept for the nonconvex
nondifferentiable game, named (restricted) B-Quasi Generalized Nash
Equilibrium (B-QGNE). Roughly speaking a B-QGNE is a solution of the
first order aggregated stationarity conditions (based on directional
derivatives) of the players' optimization problems. Aiming to devise
distributed algorithms computing a (B-)QGNE, we establish a connection
between (a subclass of) such equilibria and the stationary solutions
of a suitably defined differentiable DC program with side constraints,
for which we can successfully use the DC framework developed in the
first part of the paper. To the best of our knowledge, this is the
first attempt toward a rigorous characterization of the nondifferentiability
issue in secrecy capacity multiuser resource allocation problems.
Numerical experiments show that the proposed \emph{distributed} algorithms
yield sum-secrecy-rates that are better than those achievable by \emph{centralized}
schemes (attempting to compute stationary solutions of the DC program),
and comparable to those achievable by computationally expensive techniques
attempting to obtain globally optimal solutions.

\subsubsection{MIMO Cognitive Radio design}

We consider the sum-rate maximization of CR MIMO systems, subject
to coupling interference constraints. Special cases of such a nonconvex
problem have been widely studied in the literature. The analysis is
mainly limited to local interference constraints (see \cite{SchmidtShiBerryHonigUtschick-SPMag,ZhangLiangCui2010}
and references therein), with the exception of \cite{Kim2011,DallAnese2012}
where coupling constraints are considered. However the theoretical
convergence of algorithms in \cite{Kim2011,DallAnese2012} is up to
date an open problem. Since the general optimization problem is an
instance of DC programs with shared constraints, we can apply the
framework developed in the paper and obtain for the first time a class
of distributed (primal/dual-based) algorithms with probable convergence. 

In summary, the main contributions of this paper are:

\noindent $\bullet$ A novel class of \emph{distributed} decomposition
algorithms with provable convergence for multiuser DC problems with
(convex) side constraints.

\noindent $\bullet$ A novel game theoretical formulation for the
secrecy rate maximization problem in \emph{multiple} source-destination
OFDMA networks with \emph{multiple} friendly jammers, and consequent
algorithms to compute its QGNE based on a nontrivial DC reformulation
of the game.

\noindent $\bullet$ A class of \emph{provable convergent }distributed
primal/dual algorithms for the CR MIMO sum-rate maximization problem. 

The rest of this paper is organized as follows. Section~\ref{sec:problem}
introduces the proposed multiuser DC problem with coupling constraints.
Section~\ref{sub:SCA-centralized} presents a novel decomposition
technique for computing stationary solutions of the DC problem, which
is suitable for a distributed implementation. Distributed algorithms
building on primal and dual decomposition techniques are discussed
in Section \ref{sec:Distributed-Implementation}. Section~\ref{sec:application}
customizes the proposed framework to the cooperative physical layer
security game and the CR MIMO sum-rate maximization problem. Finally,
Section~\ref{sec:conclusion} draws some conclusions.\vspace{-0.2cm}

\section{Multiuser DC-Program with side constraints\label{sec:problem}}

We consider a multiuser system composed of $I$ coupled users. Each
user $i$ makes decision on his $n_{i}$-dimensional real strategy
vector $\mathbf{x}_{i}\in\mathbb{R}^{n_{i}}$, subject to some local
constraints given by the set $\mathcal{X}_{i}\subset\mathbb{R}^{n_{i}}$.
The joint strategy set is denoted by $\mathcal{X}\triangleq{\prod_{i=1}^{I}\mathcal{X}_{i}}$;
$\vect x_{-i}\triangleq(\vect x_{j})_{j\neq i}$ is the strategy vector
of all users except user $i$;   $\mathcal{X}_{-i}\triangleq {\prod_{j\neq i}\mathcal{X}_{j}}$; and $\vect x\triangleq(\vect x_{i})_{i=1}^{I}$
denotes the strategy profile of all the users. In addition to the
private constraints, there are also side constraints in the form of
a $n_{c}$-vector function $\mathbf{h}(\mathbf{x})\triangleq(h_{j}(\mathbf{x}))_{j=1}^{n_{c}}\leq\mathbf{0}$.
The system design is formulated as a DC program in the following form:\vspace{-0.3cm}
 
\begin{equation}
\hspace{-0.4cm}\begin{array}{ll}
\underset{\vect x}{\text{minimize}} & \theta(\mathbf{x})\triangleq{\displaystyle \sum_{i=1}^{I}\left(f_{i}(\vect x)-g_{i}(\vect x)\right)}\smallskip\\
\text{subject to } & \vect x_{i}\in\mathcal{X}_{i}\,\,\,\,\,\forall\, i=1,\ldots,I\,\,\text{(private constraints)}\medskip\\
 & \vect h(\vect x)\leq\vect0 \hfill\,\,\,\,\,\,\,\,\text{(coupling constraints)}.
\end{array}\label{eq:problem}
\end{equation}
\textbf{Assumptions}. We make the following blanket assumptions:

\noindent A1) The functions $f_{i}$, $g_{i}$ for $i=1,\ldots,I$,
and $h_{j}$ for $j=1,\ldots,n_{c}$, are convex and continuously
differentiable on $\mathcal{X}$;

\noindent A2) Each set $\mathcal{X}_{i}$ is (nonempty) closed and
convex;

\noindent A3) The functions $f_{i}$ and $g_{i}$ have Lipschitz continuous
gradients on $\Xi$ with constant $L_{\nabla f_{i}}$ and $L_{\nabla g_{i}}$,
respectively; where $\Xi$ denotes the convex feasible set of (\ref{eq:problem});
let $L_{\nabla\theta}\triangleq\sum_{i}L_{\nabla f_{i}}+\sum_{i}L_{\nabla g_{i}}$.

\noindent A4) The lower level set $\mathcal{L}(\vect x^{0})\triangleq\left\{ \vect x\in\Xi\,|\,\theta(\vect x)\leq\theta(\vect x^{0})\right\} $
of the objective function $\theta$ is compact for some $\vect x^{0}\in\Xi$;

\noindent A5) The convex coupling constraints are in the separable
form: $\vect h(\vect x)\triangleq\sum_{i=1}^{I}\vect h_{i}(\vect x_{i})\leq\vect0$.

The assumptions above are quite standard and are satisfied by a large
class of practical problems. For instance, A4 guarantees that (\ref{eq:problem})
has a solution even when $\Xi$ is not bounded; of course A4 is trivially
satisfied if $\Xi$ is bounded.

Instances of problem (\ref{eq:problem}) appear in many applications,
from signal processing to communications and networking; see Sec.
\ref{sec:application} for some motivating examples. Our goal is to
obtain \emph{distributed }best-response-like algorithms for the class
of problems (\ref{eq:problem}), converging to stationary solutions.
This confronts three major challenges, namely: i) the objective function
is the sum of differences of two convex functions, and thus in general
nonconvex; ii) the objective function is not separable in the users'
strategies (each function $f_{i}$ and $g_{i}$ depends on the strategy
profile $\mathbf{x}$ of all users); and iii) there are side constraints
coupling all the optimization variables. We deal with these issues
in the following sections. 

\noindent \vspace{-0.5cm}

\section{A new best-response SCA decomposition \label{sub:SCA-centralized}}

The standard SCA-based technique for DC programs applied to (\ref{eq:problem})
would suggest solving a sequence of convex subproblems whose objective
function is obtained by linearizing at the current iterate the nonconvex
part of $\theta(\mathbf{x})$, that is $\sum_{i}g_{i}(\mathbf{x})$,
while retaining the convex part $\sum_{i}f_{i}(\mathbf{x})$; see,
e.g., \cite{DC-intro-3,SCA-3}. However, because of the nonseparability
of the aforementioned terms, the resulting convexification does not
enjoy a decomposable structure; therefore such SCA techniques will
lead to centralized solution methods.

Here we introduce a new decomposition technique that does not suffer
from this drawback. To formally describe our approach, let us start
rewriting each function $f_{i}(\mathbf{x})$ as: given $\vect x^{\nu}$
and denoting $\vect x_{-i}^{\nu}\triangleq\left(\vect x_{j}^{\nu}\right)_{j\neq i}$, we have
\begin{equation}
f_{i}(\vect x)=f_{i}(\vect x_{i},\vect x_{-i}^{\nu})+\left[f_{i}(\vect x_{i},\vect x_{-i})-f_{i}(\vect x_{i},\vect x_{-i}^{\nu})\right].\label{eq:f_i_eq}
\end{equation}
We now approximate the term in brackets using a first order Taylor
expansion at $\vect x^{\nu}$, 
\[
f_{i}(\vect x_{i},\vect x_{-i})-f_{i}(\vect x_{i},\vect x_{-i}^{\nu})\approx\sum_{j\neq i}\nabla_{\vect x_{j}}f_{i}(\vect x^{\nu})^{T}(\vect x_{j}-\vect x_{j}^{\nu}),\vspace{-0.2cm}
\]
and approximate (\ref{eq:f_i_eq}) as 
\begin{equation}
f_{i}(\vect x)\approx\widetilde{{f}}_{i}(\vect x;\vect x^{\nu})\triangleq\underbrace{f_{i}(\vect x_{i},\vect x_{-i}^{\nu})}_{\text{diagonalization }}+\underbrace{\sum_{j\neq i}\nabla_{\vect x_{j}}f_{i}(\vect x^{\nu})^{T}(\vect x_{j}-\vect x_{j}^{\nu})}_{\text{off-diagonal linearization}}\label{eq:lin_plus_diag_f_i}
\end{equation}
yielding a \emph{diagonalization }plus \emph{off-diagonal linearization}
of $f_{i}(\vect x)$ at $\mathbf{x}^{\nu}$. To deal with the nonconvexity
of $-g_{i}(\vect x)$, we replace the functions $g_{i}(\vect x)$
with its linearization at $\mathbf{x}^{\nu}$:\vspace{-0.1cm} 
\begin{equation}
g_{i}(\vect x)\approx\widetilde{{g}}_{i}(\vect x;\vect x^{\nu})\triangleq\underbrace{g_{i}(\vect x^{\nu})+\sum_{j=1}^{I}\nabla_{\vect x_{j}}g_{i}(\vect x^{\nu})^{T}(\vect x_{j}-\vect x_{j}^{\nu})}_{\text{linearization}}.\vspace{-0.1cm}\label{eq:g_i_linearization}
\end{equation}
Based on (\ref{eq:lin_plus_diag_f_i}) and (\ref{eq:g_i_linearization}),
the candidate approximation of the nonconvex sum-utility $\theta(\mathbf{x})$
at $\mathbf{x}^{\nu}$ is:\vspace{-0.1cm} 
\begin{equation}
\widetilde{{\theta}}(\vect x;\vect x^{\nu})\triangleq\sum_{i=1}^{I}\left(\widetilde{{f}}_{i}(\vect x;\vect x^{\nu})-\widetilde{{g}}_{i}(\vect x;\vect x^{\nu})\right)+\sum_{i=1}^{I}\frac{\tau_{i}}{2}\|\vect x_{i}-\vect x_{i}^{\nu}\|^{2},\label{theta hat1}
\end{equation}
where we added a proximal-like regularization term with $\tau_i\geq 0$, whose numerical
benefits are well-understood; see, e.g., \cite{ConvexSumSeparable-1}.
Rearranging the terms in the above sum, it is not difficult
to see that (\ref{theta hat1}) can be equivalently rewritten as \vspace{-0.2cm}

\begin{equation}
\begin{aligned}\widetilde{{\theta}}(\vect x;\vect x^{\nu})=\sum_{i=1}^{I}\widetilde{{\theta}}_{i}(\vect x_{i};\vect x^{\nu}),\end{aligned}
\vspace{-0.3cm}\label{eq:theta_rearrang}
\end{equation}
where\vspace{-0.5cm}
 
\begin{equation}
\begin{aligned}\widetilde{{\theta}}_{i}(\vect x_{i};\vect x^{\nu})\triangleq\left[f_{i}(\vect x_{i},\vect x_{-i}^{\nu})+\sum_{j\neq i}\nabla_{\vect x_{i}}f_{j}(\vect x^{\nu})^{T}(\vect x_{i}-\vect x_{i}^{\nu})\right]\\
-\left[g_{i}(\vect x^{\nu})+\sum_{j=1}^{I}\nabla_{\vect x_{i}}g_{j}(\vect x^{\nu})^{T}(\vect x_{i}-\vect x_{i}^{\nu})\right]+\frac{\tau_{i}}{2}\|\vect x_{i}-\vect x_{i}^{\nu}\|^{2}.
\end{aligned}
\end{equation}

\textcolor{black}{Roughly speaking, the main idea behind the above
approximations is to use a proper combination of }\textcolor{black}{\emph{diagonalization}}\textcolor{black}{{}
and }\textcolor{black}{\emph{partial linearization}}\textcolor{black}{{}
on the convex functions $f_{i}(\mathbf{x})$ {[}cf. (\ref{eq:lin_plus_diag_f_i}){]}
together with a linearization of the nonconvex terms $g_{i}(\mathbf{x})$
{[}cf. (\ref{eq:g_i_linearization}){]}. The diagonalization procedure
fixes the non-separability issue in $\theta(\mathbf{x})$ while preserving
the convex part in $\theta(\mathbf{x})$, whereas the linearization
of $g_{i}(\mathbf{x})$ gets rid of the nonconvex part in $\theta(\mathbf{x})$.}
Indeed, this procedure leads to the approximation function $\widetilde{{\theta}}(\vect x;\vect x^{\nu})$
at $\mathbf{x}^{\nu}$ that is \emph{separable} in the users' variables
$\mathbf{x}_{i}$ {[}each $\widetilde{{\theta}}_{i}(\vect x_{i};\vect x^{\nu})$
depends only on $\mathbf{x}_{i}$, given $\mathbf{x}^{\nu}${]} and
is\emph{ strongly convex} in $\mathbf{x}\in\Xi$.

The proposed SCA decomposition consists then in solving iteratively
(possibly with a memory) the following sequence of (strongly) convex
optimization problems: given $\mathbf{x}^{\nu}\in\Xi$, 
\begin{equation}
\begin{aligned}\wh{\vect x}\left(\vect x^{\nu}\right)\triangleq\underset{\vect x\in\Xi}{\text{argmin}}\,\,\,\widetilde{{\theta}}(\vect x;\vect x^{\nu}).\end{aligned}
\label{eq:map}
\end{equation}
The formal description of the proposed SCA technique is given in Algorithm
\ref{alg:outer loop}. Note that in Step 3 of the algorithm
we allow a memory in the update of the iterate $\mathbf{x}^{\nu}$
in the form of a convex combination via $\gamma^{\nu}\in(0,1]$ (this
guarantees $\mathbf{x}^{\nu+1}\in\Xi$). \vspace{-0.2cm}

\begin{algorithm}[h]
\textbf{Data}: $\boldsymbol{{\tau}}\triangleq(\tau_{i})_{i=1}^{I}\geq \mathbf{0},\{\gamma^{\nu}\}>0$
and $\vect x^{0}\in\Xi$. Set $\nu=0$.

$(\texttt{S.1})$: If $\vect x^{\nu}$ satisfies a termination criterion,
STOP;

$(\texttt{S.2})$: Compute $\wh{\vect x}\left(\vect x^{\nu}\right)$
{[}cf. (\ref{eq:map}){]};

$(\texttt{S.3})$: Set $\vect x^{\nu+1}\triangleq\vect x^{\nu}+\gamma^{\nu}\left(\,\wh{\vect x}\left(\vect x^{\nu}\right)-\vect x^{\nu}\,\right)$;

$(\texttt{S.4})$: $\nu\leftarrow\nu+1$ and go to $(\texttt{S.1})$.

\caption{\hspace{-3pt}\textbf{:} \label{alg:outer loop}SCA Algorithm for
the DC program (\ref{eq:problem})}
\end{algorithm}

\vspace{-0.3cm}

The  convergence  of the algorithm is studied in the next theorem, where  $c_{f_{i}}(\vect x_{-i})\geq 0$ in (\ref{eq:descent_constant-1}) is the largest constant such that $(\vect z_{i}-\vect w_{i})^T (\nabla_{\vect x_{i}} f_i (\vect z_{i},\vect x_{-i})-\nabla_{\vect x_{i}} f_i (\vect w_{i},\vect x_{-i}))\geq c_{f_{i}}(\vect x_{-i}) \| \vect z_{i}-\vect w_{i} \|^2,$ for all $\vect z_{i}, \vect w_{i}\in \mathcal{X}_i$ and $\vect x_{-i}\in \mathcal{X}_{-i}$.\vspace{-0.1cm} 

\begin{theorem}\label{prop:convergence} Given the DC program (\ref{eq:problem})
under A1-A4, suppose that $\boldsymbol{{\tau}}\triangleq(\tau_{i})_{i=1}^{I}$
and $\{\gamma^{\nu}\}$ are chosen so that one of the two following
conditions are satisfied:

\noindent \emph{(a) Constant step-size rule:} 
\begin{equation}
\gamma^{\nu}=\gamma\in(0,1]\,\,\,\,\forall\,\nu\geq0\,\,\,\,\text{ and }\,\,\,\,2\, c_{\boldsymbol{{\tau}}}>\gamma L_{\nabla\theta}.\label{eq:conv_const_ss}
\end{equation}
with\vspace{-0.1cm}
\begin{equation}
c_{\boldsymbol{{\tau}}}\triangleq\min_{i=1,\ldots,I}\left\{ \tau_{i}+\inf_{\vect z_{-i}\in\mathcal{X}_{-i}}c_{f_{i}}(\vect z_{-i})\right\}.\label{eq:descent_constant-1}
\end{equation}

\noindent \emph{(b) Diminishing step-size rule:}  
\begin{equation}
\begin{aligned}c_{\boldsymbol{{\tau}}}>0,\,\,\,\gamma^{\nu}\in(0,1],\,\,\gamma^{\nu}\rightarrow0,\,\,\mbox{and}\,\,\,{\textstyle \sum_{\nu=0}^{\infty}\gamma^{\nu}=+\infty.}\end{aligned}
\label{eq:conv_dim_ss}
\end{equation}

Then, the sequence $\{\vect x^{\nu}\}$ generated by Algorithm \ref{alg:outer loop}
converges in a finite number of iterations to a stationary point of
(\ref{eq:problem}) or every limit point of the sequence (at least
one such point exists) is a stationary solution of (\ref{eq:problem}).
\end{theorem}\vspace{-0.2cm}
 
\begin{IEEEproof}
See Appendix~\ref{sec:Appendix:-Theorem_convergence}. 
\end{IEEEproof}
\begin{remark}[On Algorithm \ref{alg:outer loop}]\rm The algorithm
implements a novel SCA decomposition technique: at each iteration
$\nu$, a separable (strongly) convex function $\widetilde{{\theta}}(\vect x;\vect x^{\nu})$
is minimized over the convex set $\Xi$. The main difference from the
classical SCA techniques (e.g., \cite{DC-intro-1,DC-intro-3,SCA-3})
is that the approximation function $\widetilde{{\theta}}(\vect x;\vect x^{\nu})$
is separable across the users. In Section \ref{sec:Distributed-Implementation},
we show that such a structure leads naturally to a distributed implementation
of Algorithm \ref{alg:outer loop}. A practical termination criterion
in Step 1 is to stop the iterates when $|{\theta}(\vect x^{\nu})-{\theta}(\vect x^{\nu-1})|\leq\delta$,
where $\delta$ is a prescribed accuracy. Finally, it is reasonable to expect the algorithm to  perform better than classical gradient
algorithms applied directly to (\ref{eq:problem}) at the cost of
no extra signaling, because the structure of the objective function
${\theta}(\vect x)$ is better explored. \end{remark}\vspace{-0.4cm}

\begin{remark}[On the choice of the free parameters]\rm Theorem \ref{prop:convergence}
offers some flexibility in the choice of ${\boldsymbol{\tau}}$ and
$\gamma^{\nu}$, while guaranteeing convergence of Algorithm \ref{alg:outer loop},
which makes it applicable in a variety of scenarios. More specifically,
one can use a constant or a diminishing rule for the step-size $\gamma^{\nu}$.

\noindent \textbf{\emph{Constant step-size rule:}} This rule resembles
analogous constant step-size rules in gradient algorithms: convergence
is guaranteed either under ``sufficiently'' small step-size $\gamma$
(given $\boldsymbol{{\tau}}$) or ``sufficiently'' large $\tau_{i}$'s
(given $\gamma\in(0,1]$), such that (\ref{eq:conv_const_ss}) is
satisfied. A special case that is worth mentioning is: $\gamma=1$
and \textcolor{black}{$2\, c_{\boldsymbol{{\tau}}}>\gamma L_{\nabla\theta}$,}
which leads to a SCA-based iterate with no memory: given $\vect x^{\nu}\in\Xi$,
$\vect x^{\nu+1}\,\triangleq\,\wh{\vect x}\left(\vect x^{\nu}\right)$.
In this particular case, we can relax a bit the convergence condition
(\ref{eq:conv_const_ss}) and require the Lipschitz continuity of
the gradients of $f_{i}$ only. The result is stated next (see attached
material for the proof).\vspace{-0.1cm}

\begin{theorem}\label{prop:convergence no memory} Let\emph{ $\{\mathbf{x}^{\nu}\}$}
be the sequence generated by Algorithm \ref{alg:outer loop}, in the
setting of Theorem \ref{prop:convergence} where however we relax
A3 by assuming that $f_{i}$ have Lipschitz gradients with constants
$L_{\nabla f_{i}}$. Suppose that\emph{ }$\gamma^{\nu}=1$ and $\boldsymbol{\tau}>\mathbf{0}$
is such that \textcolor{black}{$\tau^{\min}\triangleq\min_{i}\tau_{i}>2\sum_{i=1}^{I}L_{\nabla f_{i}}$};
then, the conclusions of Theorem \ref{prop:convergence} hold.\end{theorem}\vspace{-0.1cm}

\noindent \textbf{\emph{Diminishing step-size rule:}} The application
of a constant step-size rule requires the knowledge of the Lipschitz
constants $L_{\nabla f_{i}}$ and $L_{\nabla g_{i}}$, which may not
be available. One can use a (conservative) estimate of such values
(e.g., using upper bounds), but in practice this generally leads to
``large'' values of $c_{\boldsymbol{{\tau}}}$ satisfying (\ref{eq:conv_const_ss}),
which reasonably slows down  the algorithm. In all
these situations, a valid alternative is to use a diminishing step-size
in the form (\ref{eq:conv_dim_ss}); examples of such rules are \cite{DDPA-2}:
given $\gamma^{0}=1$,\vspace{-0.1cm} 
\begin{eqnarray}
\text{Rule 1:}\quad & \gamma^{\nu}=\gamma^{\nu-1}\left(1-\epsilon\,\gamma^{\nu-1}\right), & \,\nu=1,\ldots,\label{eq:step-size_diminishing}\\
\text{Rule 2:}\quad & \gamma^{\nu}=\frac{\gamma^{\nu-1}+\beta_{1}}{1+\beta_{2}\nu}, & \,\nu=1,\ldots,\label{eq:step-size_diminishing_2}
\end{eqnarray}
where $\epsilon\in(0,1)$ and $\beta_{1},\beta_{2}\in(0,1)$ are given
constants such that $\beta_{1}\leq\beta_{2}$. \end{remark}\vspace{-0.4cm}

\subsection{Inexact implementation of $\wh{\vect x}\left(\vect x^{\nu}\right)$\label{sub:Inexact-SCA}}

We can reduce the computational effort of Algorithm \ref{alg:outer loop}
by allowing inexact computations of the solution $\wh{\vect x}\left(\vect x^{\nu}\right)$.
The convergence of the resulting algorithm is still guaranteed under more
stringent conditions on the step-size and some requirements on the
computational errors. The inexact version of Algorithm \ref{alg:outer loop}
is formally described in Algorithm \ref{alg:outer loop_inexact} below,
where Step 2 of Algorithm \ref{alg:outer loop}, the exact computation
of $\wh{\vect x}\left(\vect x^{\nu}\right)$, is replaced now with
its inexact version, that is find a $\vect z^{\nu}$ such that $\|\vect z^{\nu}-\wh{\vect x}\left(\vect x^{\nu}\right)\|\leq\varepsilon^{\nu}$,
with $\varepsilon^{\nu}$ being the accuracy in the computation of
$\wh{\vect x}\left(\vect x^{\nu}\right)$ at iteration $\nu$. \vspace{-0.2cm}

\begin{algorithm}[h]
\textbf{Data}: $\boldsymbol{{\tau}}\geq\mathbf{0},\{\gamma^{\nu}\}>0,\{\varepsilon^{\nu}\}\downarrow0$,
and $\vect x^{0}\in\Xi$. Set $\nu=0$.

$(\texttt{S.1})$: If $\vect x^{\nu}$ satisfies a termination criterion,
STOP;

$(\texttt{S.2})$: Find a $\vect z^{\nu}$ such that $\|\vect z^{\nu}-\wh{\vect x}\left(\vect x^{\nu}\right)\|\leq\varepsilon^{\nu}$;

$(\texttt{S.3})$: Set $\vect x^{\nu+1}\triangleq\vect x^{\nu}+\gamma^{\nu}\left(\,\vect z^{\nu}-\vect x^{\nu}\,\right)$;

$(\texttt{S.4})$: $\nu\leftarrow\nu+1$ and go to $(\texttt{S.1})$.

\caption{\hspace{-3pt}\textbf{:} \label{alg:outer loop_inexact}Inexact version
of Algorithm \ref{alg:outer loop}}
\end{algorithm}

\vspace{-0.2cm}

Convergence is guaranteed if the error sequence $\{\varepsilon^{\nu}\}$
and the step-size $\{\gamma^{\nu}\}$ are properly chosen, as stated
in Theorem \ref{prop:convergence with error}. The proof of this result
is based on the application of Proposition~\ref{prop:map properties}
(cf. Appendix \ref{sec:Appendix:-Theorem_convergence}) and \cite[Th. 4]{DDPA-2},
and is omitted.\begin{theorem} \label{prop:convergence with error}
Let\emph{ $\{\mathbf{x}^{\nu}\}$} be the sequence generated by Algorithm
\ref{alg:outer loop_inexact} in the setting of Theorem \ref{prop:convergence},
where however A4 is strengthened by assuming that $\theta$ is coercive
on $\Xi$. Suppose that $\{\gamma^{\nu}\}$ and $\{\varepsilon^{\nu}\}$
are chosen so that the following conditions are satisfied: i) $\gamma^{\nu}\in(0,1]$;
ii) $\gamma^{\nu}\rightarrow0$; iii) ${\textstyle {\sum_{\nu=0}^{\infty}\gamma^{\nu}}=+\infty}$;
iv) $\sum_{\nu=0}^{\infty}\left(\gamma^{\nu}\right)^{2}<+\infty$;
and v) ${\sum_{\nu=0}^{\infty}\varepsilon^{\nu}\,\gamma^{\nu}<+\infty}$.
Then, conclusions of Theorem \ref{prop:convergence} hold.\end{theorem}\vspace{-0.2cm}

Note that the steps-size rule in (\ref{eq:step-size_diminishing})
satisfies the square summability condition in Theorem \ref{prop:convergence with error}.
As expected, in the presence of errors, convergence of Algorithm \ref{alg:outer loop_inexact}
is guaranteed if $\varepsilon^{\nu}\rightarrow0$, meaning that the
sequence of approximated problems (\ref{eq:map}) is solved with increasing
accuracy. Note that Theorem \ref{prop:convergence with error}v) imposes
also a constraint on the rate by which $\varepsilon^{\nu}$ goes to
zero, which depends also on $\{\gamma^{\nu}\}$. An example of an error
sequence satisfying condition v) is $\varepsilon^{\nu}\leq\alpha\,\gamma^{\nu}$,
where $\alpha$ is any finite positive constant. Interestingly, such
a condition can be enforced in Algorithm \ref{alg:outer loop_inexact}
using classical error bound results in convex analysis; see, e.g.,
\cite[Ch. 6]{Facchinei-Pang_FVI03}. Two examples of error bounds
are given in Lemma \ref{Lemma_error_bound} below, where we introduced
the following residual quantities:\vspace{-0.3cm}
 
\begin{eqnarray*}
r_{\Xi}(\mathbf{z};\mathbf{x}^{\nu}) & \triangleq & \left\Vert \Pi_{\mathcal{N}(\mathbf{z},\Xi)}(-\nabla_{\mathbf{x}}\widetilde{{\theta}}(\mathbf{z};\mathbf{x}^{\nu}))+\nabla_{\mathbf{x}}\widetilde{{\theta}}(\mathbf{z};\mathbf{x}^{\nu})\right\Vert \\
l_{\Xi}(\mathbf{z};\mathbf{x}^{\nu}) & \triangleq & \left\Vert \mathbf{z}-\Pi_{\Xi}\left(\mathbf{z}-\nabla_{\mathbf{x}}\widetilde{{\theta}}(\mathbf{z};\mathbf{x}^{\nu})\right)\right\Vert \vspace{-0.1cm}
\end{eqnarray*}
with $\Pi_{\mathcal{N}(\mathbf{z},\Xi)}(\mathbf{y})$ {[}resp. $\Pi_{\Xi}(\mathbf{y})${]}
denoting the Euclidean projection of $\mathbf{y}$ onto the normal
cone $\mathcal{N}(\mathbf{z},\Xi)\triangleq\{\mathbf{y}\in\mathbb{R}^{n}:\mathbf{y}^{T}(\mathbf{x}-\mathbf{z})\leq0,\,\forall\mathbf{x}\in\Xi\}$
(resp. $\Xi$). The proof of Lemma \ref{Lemma_error_bound} is similar
to that of \cite[Prop. 6.3.1, Prop. 6.3.7]{Facchinei-Pang_FVI03}
and is omitted.%
\footnote{As a technical note, referring to \cite[Prop. 6.3.1, Prop. 6.3.7]{Facchinei-Pang_FVI03}
and using the same notation therein, we just observe here that the
proof of the aforementioned propositions can be easily extended to
the case of non-Cartesian set $K$ but strongly monotone VI functions
$\mathbf{F}$. %
}\vspace{-0.1cm}

\begin{lemma}\label{Lemma_error_bound} Given the optimization problem
(\ref{eq:map}) under assumptions A1-A4, the following hold:

\noindent (a): A (finite) constant $\xi_{1}>0$ exists such that\vspace{-0.2cm}
 
\[
\|\mathbf{z}-\widehat{\mathbf{x}}\left(\mathbf{x}^{\nu}\right)\|\leq\xi_{1}\, r_{\Xi}(\mathbf{z};\mathbf{x}^{\nu}),\quad\mathbf{z}\in\Xi;
\]
(b): A (finite) constant $\xi_{2}>0$ exists such that\vspace{-0.1cm}
 
\[
\|\mathbf{z}-\widehat{\mathbf{x}}\left(\mathbf{x}^{\nu}\right)\|\leq\xi_{2}\, l_{\Xi}(\mathbf{z};\mathbf{x}^{\nu}),\quad\mathbf{z}\in\mathbb{R}^{n}.
\]
\end{lemma}

Note that, for a non-polyhedral finitely representable (convex) set
$\Xi$ and $\mathbf{z}$, the computation of $l_{\Xi}(\mathbf{z};\mathbf{x}^{\nu})$
amounts to solving a convex optimization problem; whereas in the same
case if $\mathbf{z}\in\Xi$ satisfies a suitable Constraint Qualification
(CQ), the computation of $r_{\Xi}(\mathbf{z};\mathbf{x}^{\nu})$ reduces
to solving a convex \emph{quadratic} program. Thus $r_{\Xi}(\mathbf{z};\mathbf{x}^{\nu})$
is computationally easier than $l_{\Xi}(\mathbf{z};\mathbf{x}^{\nu})$
to be obtained but, contrary to $l_{\Xi}(\mathbf{z};\mathbf{x}^{\nu})$,
it can be used only to test vectors $\mathbf{z}$ belonging to $\Xi$.
Using the error bounds in Lemma \ref{Lemma_error_bound} and given
a step-size rule $\{\gamma^{\nu}\}$, the termination criterion in
Step 2 of the algorithm becomes then $r_{\Xi}(\mathbf{z};\mathbf{x}^{\nu})\leq\widetilde{{\alpha}}\,\gamma^{\nu}$
or $l_{\Xi}(\mathbf{z};\mathbf{x}^{\nu})\leq\widetilde{{\alpha}}\,\gamma^{\nu}$,
for some $\widetilde{{\alpha}}>0$.\vspace{-0.3cm}

\subsection{Generalizations}

So far we have restricted our attention to optimization problems in
the DC form. However, it is worth mentioning that the proposed analysis
and resulting algorithms (also those introduced in the forthcoming
sections) can be readily extended to other sum-utility functions not
necessarily in the DC form, such as $\widehat{{\theta}}(\mathbf{x})\triangleq\sum_{\ell\in\mathcal{I}_{f}}{f}_{\ell}(\vect x)$,
where the set $I_{f}\triangleq\{1,\ldots,I_{f}\}$ may be different
from the set of users $\{1,\ldots,I\}$, and each function ${f}_{\ell}$
is not necessarily expressed as the difference of two convex functions.
It is not difficult to show that in such a case the candidate approximation
function $\widetilde{{\theta}}(\vect x;\vect x^{\nu})$ still has
the form in (\ref{eq:theta_rearrang}), where each $\widetilde{{\theta}}_{i}(\vect x;\vect x^{\nu})$
is now given by 
\begin{align*}
\widetilde{{\theta}}_{i}(\vect x;\vect x^{\nu}) & \triangleq\sum_{j\in\mathcal{C}_{i}}f_{j}(\mathbf{x}_{i},\,\mathbf{x}_{-i}^{\nu})+\sum_{j\notin\mathcal{C}_{i}}\nabla_{\mathbf{x}_{i}}f_{j}(\mathbf{x}^{\nu})^{T}(\mathbf{x}_{i}-\mathbf{x}_{i}^{\nu})\\
 & \quad+\frac{\tau_{i}}{2}\|\vect x_{i}-\vect x_{i}^{\nu}\|^{2}
\end{align*}
where $\mathcal{C}_{i}$ is any subset of $\mathcal{S}_{i}\subseteq\mathcal{I}_{f}$,
with $\mathcal{S}_{i}\triangleq\left\{ j\in\mathcal{I}_{f}\!:\! f_{j}(\bullet,\mathbf{x}_{-i})\,\mbox{is convex on }\mathcal{X}_{i},\forall\mathbf{x}_{-i}\in\mathcal{X}_{-i}\right\} $.
In $\widetilde{{\theta}}_{i}(\vect x;\vect x^{\nu})$ each user linearizes
only the functions outside $\mathcal{C}_{i}$ while preserving the
convex part of the sum-utility. The choice of $\mathcal{C}_{i}\subseteq\mathcal{S}_{i}$
is a degree of freedom useful to explore the tradeoff between signaling
and convergence speed. We omit further details because of space limitation.\vspace{-0.2cm}

\section{Distributed Implementation \label{sec:Distributed-Implementation}}

In general, the implementation of Algorithms \ref{alg:outer loop}
and \ref{alg:outer loop_inexact} requires a coordination among the
users; the amount of network signaling depends on the specific application
under consideration. To alleviate the communication overhead of a
centralized implementation, it is desirable to obtain a decentralized
version of these schemes. Interestingly, the separability structure
of the approximation function $\widetilde{{\theta}}(\mathbf{x};\mathbf{x}^{\nu})$
resulting from the proposed convexification method (cf. Sec. \ref{sub:SCA-centralized})
as well as that of the coupling constraints lends itself to a parallel
decomposition of the subproblems (\ref{eq:map}) across the users
in the primal or dual domain. The proposed distributed implementations
of Step 2 of Algorithms \ref{alg:outer loop} (and Algorithm \ref{alg:outer loop_inexact})
are described in the next two subsections. \vspace{-0.4cm}

\subsection{Distributed dual-decomposition based algorithms\label{sub:dual-decomposition}}

The subproblems (\ref{eq:map}) can be solved in a distributed way
if the side constraints $\mathbf{h}(\mathbf{x})\leq\mathbf{0}$ are
dualized (under zero-duality gap). The dual problem associated with
each (\ref{eq:map}) is: given $\vect x^{\nu}\in\Xi$,\vspace{-0.1cm}
\begin{equation}
\begin{aligned}\begin{aligned}\underset{\boldsymbol{\lambda}\geq\vect0}{\text{maximize}}\,\left\{ d(\boldsymbol{\lambda};\vect x^{\nu})\triangleq\underset{\vect x\in\mathcal{X}}{\text{minimize}}\,\left\{ \widetilde{\theta}(\vect x;\vect x^{\nu})+\boldsymbol{\lambda}^{T}\vect h(\vect x)\right\} \right\} .\end{aligned}
\end{aligned}
\label{eq:dual function}
\end{equation}
Note that, under A1-A4, the inner minimization in (\ref{eq:dual function})
has a unique solution, which will be denoted by $\widehat{{\mathbf{x}}}(\boldsymbol{\lambda};\mathbf{x}^{\nu})\triangleq(\widehat{{\mathbf{x}}}_{i}(\boldsymbol{\lambda};\mathbf{x}^{\nu}))_{i=1}^{I}$,
that is\vspace{-0.1cm} 
\begin{equation}
\widehat{{\mathbf{x}}}_{i}(\boldsymbol{\lambda};\mathbf{x}^{\nu})\triangleq\underset{\vect x_{i}\in\mathcal{X}_{i}}{\text{{arg}}\text{min}}\,\left\{ \widetilde{\theta}_{i}(\vect x_{i};\vect x^{\nu})+\boldsymbol{\lambda}^{T}\vect h_{i}(\vect x_{i})\right\} .\vspace{-0.1cm}\label{eq:best-response_lagrangian}
\end{equation}
Before proceeding, let us introduce the following assumptions.

\noindent \textbf{Assumptions} \textbf{A5-A6}: A5) The side constraint
vector function $\mathbf{h}(\bullet)$ is Lipschitz continuous on
$\mathcal{X}$, with constant $L_{\mathbf{h}}$;

\noindent A6) For each subproblem (\ref{eq:map}), there is zero-duality
gap, \textcolor{black}{and the dual problem (\ref{eq:dual function})
has a non-empty solution set. }

We emphasize that the above conditions are generally satisfied by
many practical problems of interest. For example, we have zero duality
gap if some CQ is satisfied, e.g., (generalized) Slater's CQ, or the
feasible set $\Xi$ is a polyhedron.

The next lemma summarizes some desirable properties of the dual function
$d(\boldsymbol{\lambda};\vect x^{\nu})$, which are instrumental to
prove convergence of dual schemes.\vspace{-0.1cm}

\begin{lemma} \label{prop:suf cond inner conv} Under A1-A4 we have
the following:

\noindent (a): $d(\boldsymbol{\lambda};\vect x^{\nu})$ is differentiable
on $\mathbb{R}_{+}^{n_{c}}$, with gradient\vspace{-0.1cm} 
\begin{equation}
\nabla_{\boldsymbol{{\lambda}}}d(\boldsymbol{\lambda};\vect x^{\nu})=\vect h\left(\widehat{{\mathbf{x}}}(\boldsymbol{\lambda};\mathbf{x}^{\nu})\right)=\sum_{i}\vect h_{i}\left(\widehat{{\mathbf{x}}}_{i}(\boldsymbol{\lambda};\mathbf{x}^{\nu})\right).\vspace{-0.2cm}\label{eq:gradient_dual}
\end{equation}

\noindent (b): If, in addition, A5 holds, then $\nabla_{\boldsymbol{{\lambda}}}d(\boldsymbol{\lambda};\vect x^{\nu})$
is Lipschitz continuous on $\mathbb{R}_{+}^{n_{c}}$ with constant
$L_{\nabla d}\triangleq L_{\mathbf{h}}^{2}\,\sqrt{{n_{c}}}/c_{\boldsymbol{{\tau}}}$.
\end{lemma}\vspace{-0.1cm}\begin{proof}See attached material.\end{proof}\smallskip

The dual-problem can be solved, e.g., using well-known gradient algorithms
\cite{bertsekas2003convex}; an instance is given in Algorithm \ref{alg:Dual-Decomposition},
whose convergence is stated in Theorem \ref{thm:dual_convergence}
(whose proof follows from Lemma \ref{prop:suf cond inner conv} and
standard convergence results of gradient projection algorithms \cite{bertsekas2003convex}).
In (\ref{eq:sum-rate-dual_update-1}) $[\bullet]_{+}$ denotes the
Euclidean projection onto $\mathbb{R}_{+}$, i.e., $[x]_{+}\triangleq\max(0,x)$.
\vspace{-0.1cm}

\begin{algorithm}[H]
\textbf{Data}: $\boldsymbol{\lambda}^{0}\geq\mathbf{0}$, $\mathbf{x}^{\nu}$,
$\{\alpha^{t}\}>0$; set $t=0$.

(\texttt{S.2a}): If $\boldsymbol{\lambda}^{t}$ satisfies a suitable
termination criterion: STOP.

(\texttt{S.2b}): The users solve in parallel (\ref{eq:best-response_lagrangian}):
for all $i=1,\dots,I$, compute $\widehat{{\mathbf{x}}}_{i}(\boldsymbol{\lambda}^{t};\mathbf{x}^{\nu})$

(\texttt{S.2c}): Update $\boldsymbol{\lambda}$ according to 
\begin{equation}
\boldsymbol{\lambda}^{t+1}\triangleq\left[\boldsymbol{\lambda}^{t}+\alpha^{t}\,\sum_{i=1}^{I}\vect h_{i}\left(\widehat{{\mathbf{x}}}_{i}(\boldsymbol{\lambda}^{t};\mathbf{x}^{\nu})\right)\right]_{+}.\label{eq:sum-rate-dual_update-1}
\end{equation}

(\texttt{S.2d}): $t\leftarrow t+1$ and go back to (\texttt{S.2a}).

\caption{\hspace{-3pt}\textbf{: \label{alg:Dual-Decomposition}}Dual-based
Distributed Implementation of Step 2 of Algorithm \ref{alg:outer loop} }
\end{algorithm}
\vspace{-0.3cm}

\begin{theorem}\label{thm:dual_convergence} Given the DC program
(\ref{eq:problem}) under A1-A4, suppose that one of the two following
conditions are satisfied:

\noindent (a)\emph{: }A5 holds and $\{\alpha^{t}\}$ is chosen such
that $0<\alpha^{t}=\alpha^{\max}<2/L_{\nabla d}$, for all $t\geq0$;

\noindent (b)\emph{: }$\nabla_{\boldsymbol{{\lambda}}}d(\boldsymbol{\bullet};\vect x^{\nu})$
is uniformly bounded on\emph{ }$\mathbb{R}_{+}^{n_{c}}$, and $\{\alpha^{t}\}$
is chosen such that $\alpha^{t}>0$, $\alpha^{t}\rightarrow0$, $\sum_{t}\alpha^{t}=\infty$,
and $\sum_{t}(\alpha^{t})^{2}<\infty$.

Then, the sequence $\left\{ \boldsymbol{\lambda}^{t}\right\} $ generated
by Algorithm \ref{alg:Dual-Decomposition} converges to a solution
of (\ref{eq:dual function})\emph{.} If, in addition, A6 holds, the
sequence $\{\widehat{{\mathbf{x}}}(\boldsymbol{\lambda}^{t};\mathbf{x}^{\nu})\}$
converges to the unique solution of (\ref{eq:map}). \end{theorem}\vspace{-0.2cm}

\begin{remark}[On the distributed implementation]\rm The distributed
implementation of Algorithms \ref{alg:outer loop} and \ref{alg:outer loop_inexact}
based on Algorithm \ref{alg:Dual-Decomposition} leads to a double-loop
scheme with communication between the two loops: given the current
value of the multipliers $\boldsymbol{{\lambda}}^{t}$, the users
can solve in a distributed way their subproblems (\ref{eq:best-response_lagrangian});
once the new value $\widehat{{\mathbf{x}}}(\boldsymbol{\lambda}^{t};\mathbf{x}^{\nu})$
is available, the multipliers are updated according to (\ref{eq:sum-rate-dual_update-1}).
Note that when $n_{c}=1$ (i.e., only one shared constraint), the
update in (\ref{eq:sum-rate-dual_update-1}) can be replaced by a
bisection search, which generally converges quite fast. When $n_{c}>1$,
the potential slow convergence of gradient updates (\ref{eq:sum-rate-dual_update-1})
can be alleviated using accelerated gradient-based update; see, e.g.,
\cite{Nesterov04}. 
Note also that the size of the dual problem (the dimension of $\boldsymbol{\lambda}$)
is equal to $n_{c}$ (the number of shared constraints), which makes
Algorithm\emph{ }\ref{alg:Dual-Decomposition} scalable in the number
of users.

As far as the communication overhead needed to implement the proposed
scheme is concerned, the signaling among the users is in the form
of message passing and, of course, is problem dependent; see Sec. \ref{sec:application}
for specific examples. When the networks has a cluster-head, the update
of the multipliers can be performed at the cluster, and then broadcast
to the users. In fully decentralized networks, the update of $\boldsymbol{{\lambda}}$
can be done by the users themselves, by running consensus based algorithms
to locally estimate $\sum_{i}\vect h_{i}\left(\widehat{{\mathbf{x}}}_{i}(\boldsymbol{\lambda}^{t};\mathbf{x}^{\nu})\right)$.
This in general requires a limited signaling exchange among neighboring
nodes. \end{remark}\vspace{-0.4cm}

\subsection{Distributed implementation via primal-decomposition\label{sub:Primal-decomposition}}

Algorithm \ref{alg:Dual-Decomposition} is based on the relaxation
of the side constraints into the Lagrangian, resulting in general
in a violation of these constraints during the intermediate iterates.
In some applications, this may prevent the on-line implementation
of the algorithm. In this section, we propose a distributed scheme
which does not suffer from this issue; we cope with side constraints
using a primal decomposition technique.

Introducing the slack variables $\mathbf{t}\triangleq(\mathbf{t}_{i})_{i=1}^{I}$,
with each $\mathbf{t}_{i}\in\mathbb{R}^{n_{c}}$, (\ref{eq:map})
can be rewritten as\vspace{-0.3cm}
 
\begin{equation}
\begin{array}{cl}
\underset{(\mathbf{x}_{i},\mathbf{t}_{i})_{i=1}^{I}}{\textrm{minimize}} & \sum_{i=1}^{I}\widetilde{{\theta}}_{i}(\vect x_{i};\vect x^{\nu}),\\
\textrm{subject to} & \mathbf{x}_{i}\in\mathcal{X}_{i},\quad \quad \,\,\,   \forall i=1,\ldots,I,\\
 & \mathbf{h}_{i}(\mathbf{x}_{i})\leq\mathbf{t}_{i},\quad\forall i=1,\ldots,I,\\
 & \sum_{i=1}^{I}\mathbf{t}_{i}\leq\mathbf{0},
\end{array}\label{eq:primal_reformulation}
\end{equation}
When $\mathbf{t}=(\boldsymbol{t}_{i})_{i=1}^{I}$ is fixed, (\ref{eq:primal_reformulation})
can be decoupled across the users: for each $i=1,\ldots,I$, solve
\begin{equation}
\begin{array}{cl}
\underset{\mathbf{x}_{i}}{\textrm{minimize}} & \widetilde{{\theta}}_{i}(\vect x_{i};\vect x^{\nu}),\\
\textrm{subject to} & \mathbf{x}_{i}\in\mathcal{X}_{i},\\
 & \mathbf{h}_{i}(\mathbf{x}_{i})\overset{\boldsymbol{{\mu}}_{i}(\mathbf{t}_{i};\mathbf{x}^{\nu})}{\leq}\mathbf{t}_{i},
\end{array}\label{eq:primal_single_user}
\end{equation}
where $\boldsymbol{{\mu}}_{i}(\mathbf{t}_{i};\mathbf{x}^{\nu})$ is
the optimal Lagrange multiplier associated with the inequality constraint
$\mathbf{h}_{i}(\mathbf{x}_{i})\leq\mathbf{t}_{i}$. \textcolor{black}{Note
that the existence of $\boldsymbol{{\mu}}_{i}(\mathbf{t}_{i};\mathbf{x}^{\nu})$
is guaranteed if (\ref{eq:primal_single_user}) has zero duality gap
(e.g., when some CQ hold), but $\boldsymbol{{\mu}}_{i}(\mathbf{t}_{i};\mathbf{x}^{\nu})$
may not be unique.}\textcolor{red}{{} }Let us denote by $\mathbf{x}_{i}^{\star}(\mathbf{t}_{i};\mathbf{x}^{\nu})$
the unique solution of (\ref{eq:primal_single_user}), given $\mathbf{t}=$$(\boldsymbol{t}_{i})_{i=1}^{I}$.
The optimal slack $\mathbf{t}^{\star}\triangleq(\mathbf{t}_{i}^{\star})_{i=1}^{I}$
of the shared constraints can be found solving the so-called \emph{master}
(convex) problem (see, e.g., \cite{ConvexSumSeparable-3}): 
\begin{equation}
\begin{array}{cl}
\underset{\boldsymbol{t}}{\textrm{minimize}} & P(\boldsymbol{t};\mathbf{x}^{\nu})\triangleq\sum_{i=1}^{I}\widetilde{{\theta}}_{i}(\mathbf{x}_{i}^{\star}(\mathbf{t}_{i};\mathbf{x}^{\nu});\vect x^{\nu})\\
\textrm{subject to} & \sum_{i=1}^{I}\mathbf{t}_{i}\leq\mathbf{0}.
\end{array}\label{eq:primal_master}
\end{equation}
Due to the non-uniqueness of $\boldsymbol{{\mu}}_{i}(\mathbf{t}_{i};\mathbf{x}^{\nu})$,
the objective function in (\ref{eq:primal_master}) is nondifferentiable;
problem (\ref{eq:primal_master}) can be solved by subgradient methods.
The subgradient of $P(\boldsymbol{t};\mathbf{x}^{\nu})$ at $\mathbf{t}$
is \textcolor{black}{{} 
\[
\partial_{\mathbf{t}_{i}}P(\mathbf{t};\mathbf{x}^{\nu})=-\boldsymbol{{\mu}}_{i}(\mathbf{t}_{i};\mathbf{x}^{\nu}),\quad i=1,\ldots,I.
\]
} We refer to \cite[Prop. 8.2.6]{bertsekas2003convex} for standard
convergence results of sub-gradient projection algorithms.

\vspace{-0.1cm}

\section{Applications and Numerical Results\label{sec:application}}

The DC formulation (\ref{eq:problem}) and the consequent proposed
algorithms are general enough to encompass many problems of practical
interest in different fields. Because of the space limitation, here
we focus only on two specific applications that can be casted in the
DC-form (\ref{eq:problem}), namely: i) a new secrecy rate maximization
game; and ii) the sum-rate maximization problem over MIMO CR networks.
\vspace{-0.2cm}

\subsection{Case Study \#1: A new secrecy rate game}

\subsubsection{System model and problem formulation\label{sub:System-model}}

Consider a wireless communication system composed of $Q$ transmitter-receiver
pairs$-$the legitimate users$-J$ friendly jammers, and a single
eavesdropper. We assume OFDMA transmissions for the legitimate users
over flat-fading and quasi-static (constant within the transmission
frame) channels. We denote by $H_{qq}^{\texttt{{SD}}}$ the channel
gain of the legitimate source-destination pair $q$, by ${H}_{jq}^{\texttt{{JD}}}$
the channel gain between jammer $j$ and the receiver of user $q$,
by ${H}_{je}^{\texttt{{JE}}}$ the channel gain between the transmitter
of jammer $j$ and the receiver of the eavesdropper, and by $H_{qe}^{\texttt{{SE}}}$
the channel between the source $q$ and the eavesdropper. We assume
CSI of the eavesdropper's (cross-)channels; this is a common assumption
in PHY security literature; see, e.g., \cite{Jorswieck2010,CooperativeJamming-7,CooperativeJamming-8,CooperativeJamming-Petropulu2,GameTheoreticalModel-5,GameTheoreticalModel-6,CooperativeJamming-Ma,GameTheoreticalModel-2,li2013transmit}.
CSI on the eavesdropper's channel can be obtained when the eavesdropper
is active in the network and its transmissions can be monitored. For
instance, this happens in   networks combining multicast and unicast
transmissions wherein the users work as legitimate receiver for some
signals and eavesdroppers for others. Another scenario is a cellular
environment where the eavesdroppers are also users of the network;
in a time-division duplex system, the base station can estimate users'
channels based on the channel reciprocity.

In the setting above, we adopt the CJ paradigm: the friendly jammers
cooperate with the users by introducing a proper interference profile
``masking'' the eavesdropper. The (uniform) power allocation of
source $q$ is denoted by $p_{q}$; $p_{jq}^{\texttt{{J}}}$ is the
fraction of power of friendly jammer $j$ requested by user $q$ (allocated
by jammer $j$ over the channel used by user $q$); the power profile
allocated by all the jammers over the channel of user $q$ is ${\vect p}_{q}^{\texttt{{J}}}\triangleq(p_{jq}^{\texttt{{J}}})_{j=1}^{J}$.
Each user $q$ has power budget $p_{q}\leq P_{q}$, and likewise each
jammer $j$, that is $\sum_{q=1}^{Q}p_{jq}^{\texttt{{J}}}\leq P_{j}^{\texttt{{J}}}$,
for all $j=1,\ldots,J$. Under basic information theoretical assumptions,
the maximum achievable rate on link $q$ is\vspace{-0.1cm} 
\begin{equation}
\begin{aligned}r_{qq}\left(p_{q},{\vect p}_{q}^{\texttt{{J}}}\right)\triangleq\log\left(1+\frac{H_{qq}^{\texttt{{SD}}}\, p_{q}}{\sigma^{2}+{\sum_{j=1}^{J}{H}_{jq}^{\texttt{{JD}}}\, p_{jq}^{\texttt{{J}}}}}\right)\end{aligned}
.\vspace{-.1cm}\label{eq:R_qq}
\end{equation}
Similarly, the rate on the channel between source $q$ and the eavesdropper
is\vspace{-0.2cm} 
\begin{equation}
\begin{aligned}r_{qe}\left(p_{q},\mathbf{p}_{q}^{\texttt{{J}}}\right)\triangleq\log\left(1+\frac{H_{qe}^{\texttt{{SE}}}\, p_{q}}{\sigma^{2}+{\sum_{j=1}^{J}{H}_{je}^{\texttt{{JE}}}\, p_{jq}^{\texttt{{J}}}}}\right).\end{aligned}
\label{eq:R_q0}
\end{equation}
The secrecy rate of user $q$ is then (see, e.g., \cite{Jorswieck2010}):
\begin{equation}
r_{q}^{s}\left(p_{q},\mathbf{p}_{q}^{\texttt{{J}}}\right)\triangleq\left[\, r_{qq}(p_{q},\mathbf{p}_{q}^{\texttt{{J}}})\,-\, r_{qe}(p_{q},\mathbf{p}_{q}^{\texttt{{J}}})\,\right]_{+}.\label{eq:sc}
\end{equation}

\noindent \textbf{Problem formulation:} We formulate the system design
as a game where the legitimate users are the players who cooperate
with the jammers to maximize their own secrecy rate. More formally,
anticipating $\left({\vect p}_{r}^{\texttt{{J}}}\right)_{r\neq q}$,
each user $q$ seeks together with the jammers the tuple $\left(p_{q},{\vect p}_{q}^{\texttt{{J}}}\right)$
solving the following optimization problem:\vspace{-0.1cm} 
\begin{equation}
\!\!\mathcal{G}:\begin{aligned}\!\begin{array}{ll}
\!\!\!\!\underset{\left(p_{q},{\vect p}_{q}^{\texttt{{J}}}\right)\geq\mathbf{0}}{\text{maximize}\!\!\!}\!\!\! & r_{q}^{s}\left(p_{q},\mathbf{p}_{q}^{\texttt{{J}}}\right)\\
\!\!\!\text{subject to:} & \begin{array}[t]{c}
\vspace{-0.5cm}\\
\hspace{-0.2cm}\hspace{-0.2cm}\hspace{-0.2cm}\begin{array}[t]{l}
\left.\begin{array}{l}
p_{q}\leq P_{q},\medskip\\
\sum_{r=1}^{Q}\: p_{jr}^{\texttt{{J}}}\leq P_{j}^{\texttt{{J}}},\,\, \forall\, j=1,\ldots,J.
\end{array} \hspace{-0.05 in}\right\} \!\!\triangleq\!\mathcal{P}_{q}(\mathbf{p}_{-q}^{\texttt{{J}}})\end{array}
\end{array}
\end{array}\end{aligned}
\label{eq:SCproblem}
\end{equation}
Note that the feasible set $\mathcal{P}_{q}(\mathbf{p}_{-q}^{\texttt{{J}}})$
of (\ref{eq:SCproblem}) depends on the jammers' power profile $\mathbf{p}_{-q}^{\texttt{{J}}}\triangleq(\mathbf{p}_{r}^{\texttt{{J}}})_{r\neq q}$
allocated over the other users' channels. When needed, we will denote
each tuple $(p_{q},\mathbf{p}_{q}^{\texttt{{J}}})$ by $\mathbf{x}_{q}\triangleq(p_{q},\mathbf{p}_{q}^{\texttt{{J}}})$.
The above game whose $q$-th optimization problem is given by (\ref{eq:SCproblem})
will be termed as secrecy game $\mathcal{G}$.

The secrecy game $\mathcal{G}$ is an instance of the so-called Generalized
Nash Equilibrium problem (GNEP) with shared constraints; see, e.g.,
\cite{FacchineiKanzow07}. A solution of $\mathcal{G}$ is the Generalized
Nash Equilibrium (GNE), defined as follows. \begin{definition}[GNE]\label{Def_GNE}
A strategy profile $(p_{q}^{\star},{\vect p}_{q}^{\texttt{{J}}\star})_{q=1}^{Q}$
is a GNE of the GNEP $\mathcal{G}$ if the following holds for all
$q=1,\ldots,Q$: $(p_{q}^{\star},{\vect p}_{q}^{\texttt{{J}}\star})\in\mathcal{P}_{q}(\mathbf{p}_{-q}^{\texttt{{J}}\star})$
and\vspace{-0.1cm} 
\begin{equation}
r_{q}^{s}(p_{q}^{\star},{\vect p}_{q}^{\texttt{{J}}\star})\geq r_{q}^{s}(p_{q},{\vect p}_{q}^{\texttt{{J}}}),\quad\forall(p_{q},{\vect p}_{q}^{\texttt{{J}}})\in\mathcal{P}_{q}({\vect p}_{-q}^{\texttt{{J}}\star}).\label{eq:GNE_G}
\end{equation}
\end{definition} The (distributed) computation of a GNE of $\mathcal{G}$
is a challenging if not an impossible task, due to the following issues:
1) the non-differentiability of the player's objective functions;
2) the non-concavity of the players' objective functions; and 3) the
presence of coupling constraints. Toward the practical resolution
of $\mathcal{G}$ coping with the aforementioned issues, we follow
the three logical steps outlined next. 

\noindent $-$\emph{A smooth-game formulation}: We start dealing with
the nondifferentiability issue by introducing a smooth restricted
(still nonconcave) version of the original game $\mathcal{G}$, termed
game $\mathcal{G}^{\text{\texttt{{sm}}}}$, and establishing the connection
with $\mathcal{G}$ in terms of GNE;

\noindent $-$\emph{Relaxed equilibrium concepts}: To deal with the
nonconcavity of the players' objective functions we introduce relaxed
equilibrium concepts for both games $\mathcal{G}$ and $\mathcal{G}^{\text{\texttt{{sm}}}}$,
and establish their connections. The comparison shows that the smooth
game $\mathcal{G}^{\text{{sm}}}$ preserves (relaxed) solutions of
$\mathcal{G}$ of practical interest while just ignoring those yielding
zero secrecy rates for the players. 

\noindent $-$\emph{Algorithmic design}: We then focus on the computation
of the relaxed equilibria of $\mathcal{G}^{\text{\texttt{{sm}}}}$,
casting the problem into a multiuser DC program in the form (\ref{eq:problem}),
which can be distributively solved using the machinery introduced
in the first part of the paper.

\subsubsection{A smooth-game formulation}

We introduce a restricted smooth-game wherein the max operator in
the objective function of each player's optimization problem (\ref{eq:SCproblem})
is relaxed via linear constraints. This formulation is a direct consequence
of the following fact: $r_{qq}\left(p_{q},{\vect p}_{q}^{\texttt{{J}}}\right)\geq r_{qe}\left(p_{q},{\vect p}_{q}^{\texttt{{J}}}\right)$
if and only if either $p_{q}=0$ or\vspace{-0.2cm} 
\[
\frac{H_{qq}^{\texttt{{SD}}}}{\sigma^{2}+{\sum_{j=1}^{J}{H}_{jq}^{\texttt{{JD}}}\, p_{jq}^{\texttt{{J}}}}}\geq\frac{H_{qe}^{\texttt{{SE}}}}{\sigma^{2}+{\sum_{j=1}^{J}{H}_{je}^{\texttt{{JE}}}\, p_{jq}^{\texttt{{J}}}}}.
\]
Clearly, the latter inequality is equivalent to:\vspace{-0.2cm} 
\begin{equation}
\sum_{j=1}^{J}\left(H_{qq}^{\texttt{{SD}}}{H}_{je}^{\texttt{{JE}}}-H_{qe}^{\texttt{{SE}}}{H}_{jq}^{\texttt{{JD}}}\right)p_{jq}^{\texttt{J}}+\left(H_{qq}^{\texttt{{SD}}}-H_{qe}^{\texttt{{SE}}}\right)\sigma^{2}\geq0.\vspace{-0.1cm}\label{eq:linear_const}
\end{equation}
Note that if (\ref{eq:linear_const}) holds with equality, then $r_{q}^{s}\left(p_{q},\mathbf{p}_{q}^{\texttt{{J}}}\right)$
= 0 for any $p_{q}\geq0$. These observations lead to the following
restricted smooth game, which we call $\mathcal{G}^{\texttt{sm}}$,
where each player $q$ anticipating $\vect p_{-q}^{\texttt{J}}$ solves
the following smooth, albeit nonconcave, maximization problem:  
\begin{equation}
\!\!\mathcal{G}^{\text{\texttt{{sm}}}}\!:\!\begin{aligned} & \underset{\left(p_{q},{\vect p}_{q}^{\texttt{{J}}}\right)\geq\mathbf{0}}{\text{maximize}\,\,\,\,\,\,}\widetilde{{r_{q}^{s}}}(p_{q},\mathbf{p}_{q}^{\texttt{{J}}})\triangleq r_{qq}(p_{q},\mathbf{p}_{q}^{\texttt{{J}}})\,-\, r_{qe}(p_{q},\mathbf{p}_{q}^{\texttt{{J}}})\\
 & \text{subject to:}\,\,\,\,\,\,\begin{array}[t]{c}
\vspace{-0.4cm}\\
\hspace{-0.2cm}\hspace{-0.2cm}\hspace{-0.2cm}\begin{array}[t]{l}
\left.\begin{array}{l}
p_{q}\leq P_{q},\\
\mbox{constraint }(\ref{eq:linear_const}),\\
\sum_{r=1}^{Q}\: p_{jr}^{\texttt{{J}}}\leq P_{j}^{\texttt{{J}}},\forall\, j=1,\ldots,J,
\end{array}\!\!\!\!\right\} \!\!\triangleq\!\mathcal{P}_{q}^{\text{\texttt{{sm}}}}(\mathbf{p}_{-q}^{\texttt{{J}}})\end{array}
\end{array}
\end{aligned}
\label{SCproblem-smooth}
\end{equation}
where we denoted by $\mathcal{P}_{q}^{\text{{sm}}}(\mathbf{p}_{-q}^{\texttt{{J}}})$
the feasible set of the optimization problem. For notational convenience,
we also introduce the joint strategy set $\mathcal{P}$ defined as:
$\mathcal{P}\triangleq$\vspace{-0.1cm} 
\begin{equation}
\hspace{-0.1cm}\left\{ \!\!\!\!\begin{array}{lll}
(p_{q},{\vect p}_{q}^{\texttt{{J}}}))_{q=1}^{Q}\geq\mathbf{0}\!\!\! & \!\!:\!\! & \!\!\!\! p_{q}\leq P_{q}\,\,\mbox{and}\,\,(\ref{eq:linear_const})\mbox{ holds, }\,\forall q=1,\ldots,Q,\smallskip\\
 &  & \!\!\!{\displaystyle \!\!\sum}_{r=1}^{Q}\: p_{jr}^{\texttt{{J}}}\leq P_{j}^{\texttt{{J}}},\quad\forall\, j=1,\ldots,J
\end{array}\!\!\!\!\right\} \label{eq:joint-set}
\end{equation}

It turns out from the above discussion that, solution-wise, the main
difference between the smooth game $\mathcal{G}^{\texttt{{sm}}}$
and the original one $\mathcal{G}$ is that $\mathcal{G}^{\texttt{{sm}}}$
ignores the feasible players' strategy profiles of $\mathcal{G}$
violating (\ref{eq:linear_const}) (and thus outside $\mathcal{P}$).
But such tuples yield zero secrecy rate of the players and thus are
of little significance, since the players' goals are to attempt the
maximization of their secrecy rates. We can then focus on strategy
profiles in the set $\mathcal{P}$, without any practical loss of
optimality. The next proposition makes formal the aforementioned connection
between $\mathcal{G}$ and $\mathcal{G}^{\texttt{{sm}}}$.\vspace{-0.2cm}

\begin{proposition} \label{prop:connections-GNE} Given $\mathcal{G}$
and $\mathcal{G}^{\text{\texttt{{sm}}}}$, the following hold.

\item (a) A GNE of $\mathcal{G}$ always exists;

\item (b) A GNE of $\mathcal{G}^{\text{\texttt{{sm}}}}$ exists provided
that $\mathcal{P}\neq\emptyset$;

\item (c) $\!\![\mathcal{G}\rightarrow\mathcal{G}^{\text{\texttt{{sm}}}}]$\emph{:}
If $\vect x^{\star}$ is a GNE of $\mathcal{G}$ satisfying the constraints
(\ref{eq:linear_const}) for all $q=1,\ldots,Q$, then $\vect x^{\star}$
is a GNE of $\mathcal{G}^{\text{\texttt{{sm}}}}$.

\item (d) \textcolor{black}{$\!\![\mathcal{G}^{\text{\texttt{{sm}}}}\rightarrow\mathcal{G}]$}\textcolor{black}{\emph{:
}}\textcolor{black}{If $\vect x^{\star}$ is a GNE of $\mathcal{G}^{\text{\texttt{{sm}}}}$,
then $\vect x^{\star}$ is a GNE of $\mathcal{G}$.}\end{proposition}\vspace{-0.1cm}
 
\begin{IEEEproof}
See Appendix \ref{Proof-of-Proposition_GNE}. 
\end{IEEEproof}
The first important result stated in Proposition \ref{prop:connections-GNE}
is that the two games have a solution ($\mathcal{G}^{\texttt{{sm}}}$
under $\mathcal{P}\neq\emptyset$). Note that a sufficient condition
guaranteeing $\mathcal{P}\neq\emptyset$ is $H_{qq}^{\texttt{{SD}}}\geq H_{qe}^{\texttt{{SE}}}$
for all $q$ {[}cf. (\ref{eq:linear_const}){]}, implying that the
channel gains of the legitimate users cannot be worse than those between
the sources and the eavesdropper. For instance, this happens if the
legitimate receivers are much closer to their intended transmitters
than the eavesdropper's receiver. Conversely, if (\ref{eq:linear_const})
is violated for some $q$ (implying $\mathcal{P}=\emptyset$), there
exists no feasible user/jammer power allocation yielding a positive
secrecy rate for user $q$. Note that this is in agreement with current
results in the literature; see, e.g., \cite{GameTheoreticalModel-2,GameTheoreticalModel-6}.

Proposition \ref{prop:connections-GNE} also establishes the connection
between the GNE of $\mathcal{G}$ and $\mathcal{G}^{\texttt{{sm}}}$.
In particular, statement (d) justifies the focus on the smooth   
(and thus more affordable) game $\mathcal{G}^{\texttt{{sm}}}$ rather
than the nonsmooth $\mathcal{G}$ without any practical loss of generality.
The computation of a GNE of $\mathcal{G}^{\texttt{{sm}}}$ however
remains a difficult task, because of the nonconcavity of the players'
objective functions $\widetilde{{r_{q}^{s}}}$. To obtain practical
solution schemes, we introduce next a relaxed equilibrium concept,
whose computation can be done using the framework developed in the
first part of the paper. 

\subsubsection{Relaxed equilibrium concepts\label{sub:Relaxed-equilibrium-concepts}}

Based on the concept of B(ouligand)-derivative \cite{Facchinei-Pang_FVI03}
we define next a relaxed notion of equilibrium for the nonsmooth nonconvex
game $\mathcal{G}$.\vspace{-0.1cm}
 \begin{definition}[B-QGNE]\label{Def_BQNE} A strategy profile $\mathbf{x}^{\star}\triangleq(\mathbf{x}_{q}^{\star}\triangleq(p_{q}^{\star},{\vect p}_{q}^{\texttt{{J}}\star}))_{q=1}^{Q}$
is a B-Quasi GNE (B-QGNE) of $\mathcal{G}$ if the following holds
for all $q=1,\ldots,Q$: $(p_{q}^{\star},{\vect p}_{q}^{\texttt{{J}}\star})\in\mathcal{P}_{q}(\mathbf{p}_{-q}^{\texttt{{J}}\star})$
and\vspace{-0.2cm} 
\begin{equation}
r_{q}^{s}{'}(\mathbf{x}_{q}^{\star};\mathbf{x}_{q}-\mathbf{x}_{q}^{\star})\le0\quad\forall\mathbf{x}_{q}\in\mathcal{P}_{q}({\vect p}_{-q}^{\texttt{{J}}\star}),\label{eq:B-QGNE}
\end{equation}
where $r_{q}^{s}{'}(\mathbf{x}_{q}^{\star};\mathbf{x}_{q}-\mathbf{x}_{q}^{\star})$
denotes the directional derivative of the function $r_{q}^{s}$ at
$\mathbf{x}_{q}^{\star}$ along the direction $\mathbf{x}_{q}-\mathbf{x}_{q}^{\star}$.
\end{definition} In words, a B-QGNE is a stationary solution of the
GNEP, where the stationary concept is based on the directional derivative. Of course the B-QGNE can be defined also for the smooth game
$\mathcal{G}^{\text{{sm}}};$ in such a case, the directional derivative
$r_{q}^{s}{'}(\mathbf{x}_{q}^{\star};\mathbf{x}_{q}-\mathbf{x}_{q}^{\star})$
in (\ref{eq:B-QGNE}) reduces to $\nabla_{\mathbf{x}_{q}}\widetilde{{r_{q}^{s}}}(\mathbf{x}_{q}^{\star})^{T}(\mathbf{x}_{q}-\mathbf{x}_{q}^{\star})$,
because $\widetilde{{r_{q}^{s}}}(\bullet)$ is differentiable; thus,
in the following, we will refer to it just as QGNE of $\mathcal{G}^{\text{{sm}}}.$
Note that the B-QGNE is an instance of the Quasi-Nash Equilibrium
introduced recently in \cite{nonconvexgame,Pang-Scutari-NNConvex_PI}
to deal with the nonconvexity of the players' optimization problems.

It is worth mentioning that since the players' optimization problems
in $\mathcal{G}$ and $\mathcal{G}^{\text{{sm}}}$ have polyhedral
feasible sets, every GNE of $\mathcal{G}$ (resp. $\mathcal{G}^{\text{{sm}}}$)
is a B-QGNE (resp. QGNE), but the converse is not necessarily true.

As already observed for the GNE, the subclass of quasi-solutions of
$\mathcal{G}$ of practical interest are those associated with the
strategy profiles belonging to the set $\mathcal{P}$. We can capture
this feature introducing the concept of \emph{restricted} B-QGNE of
$\mathcal{G}$, which are B-QGNE ``over the set $\mathcal{P}$'',
ignoring thus all feasible tuples of (\ref{eq:SCproblem}) that yield
zero secrecy rates.

\begin{definition}[Restricted B-QGNE]\label{Def_restrictedBQNE}
A strategy profile $(\mathbf{x}_{q}^{\star}\triangleq(p_{q}^{\star},{\vect p}_{q}^{\texttt{{J}}\star}))_{q=1}^{Q}$
is a \emph{restricted} B-GGNE of $\mathcal{G}$ if the following holds
for all $q=1,\ldots,Q$: $(p_{q}^{\star},{\vect p}_{q}^{\texttt{{J}}\star})\in\mathcal{P}_{q}(\mathbf{p}_{-q}^{\texttt{{J}}\star})$
and 
\[
r_{q}^{s}{'}(\mathbf{x}_{q}^{\star};\mathbf{x}_{q}-\mathbf{x}_{q}^{\star})\le0,\quad\forall\mathbf{x}_{q}\in\mathcal{P}_{q}^{\texttt{sm}}({\vect p}_{-q}^{\texttt{{J}}\star}),
\]
with $\mathcal{P}_{q}^{\texttt{sm}}({\vect p}_{-q}^{\texttt{{J}}\star})$
defined in (\ref{SCproblem-smooth}).\end{definition}

A natural question now is whether the connection between the GNE of
$\mathcal{G}$ and $\mathcal{G}^{\texttt{{sm}}}$ as stated in Proposition
\ref{prop:connections-GNE} is somehow preserved also in terms of
quasi-equilibria (which in principle is not guaranteed). The answer
is stated next. \vspace{-0.1cm}

\begin{proposition} \label{prop:connections_QGNE} Given $\mathcal{G}$
and $\mathcal{G}^{\text{\texttt{{sm}}}}$, the following hold.

\item (a) A B-QGNE of $\mathcal{G}$ always exists;

\item (b) A QGNE of $\mathcal{G}^{\text{\texttt{{sm}}}}$ (resp.
restricted B-QGNE of $\mathcal{G}$) exists provided that $\mathcal{P}\neq\emptyset$;

\item (c) $[\mathcal{G}\rightarrow\mathcal{G}^{\text{\texttt{{sm}}}}]$\emph{:}
If $\vect x^{\star}$ is a (restricted) B-QGNE of $\mathcal{G}$ satisfying
(\ref{eq:linear_const}) for all $q=1,\ldots,Q$, then $\vect x^{\star}$
is a QGNE of $\mathcal{G}^{\text{\texttt{{sm}}}}$;

\item (d) $[\mathcal{G}^{\text{\texttt{{sm}}}}\rightarrow\mathcal{G}]$\emph{:}
\textcolor{black}{If $\vect x^{\star}$ is a QGNE of $\mathcal{G}^{\texttt{sm}}$
, then $\vect x^{\star}$ is a restricted B-QGNE of $\mathcal{G}$.}\end{proposition}\vspace{-0.3cm}
 
\begin{IEEEproof}
See Appendix \ref{sec:Proof-of-Proposition_QGNE}. 
\end{IEEEproof}
The above proposition paves the way to the design of numerical methods
to compute a B-QGNE of $\mathcal{G}$. Indeed, according to statement
(d), one can compute a B-QGNE of $\mathcal{G}$ (which is a solutions
of practical interest) via a QGNE of $\mathcal{G}^{\texttt{{sm}}}$.
This last task is addressed in the next subsection.

\subsubsection{Algorithmic design\label{sub:Algorithmic-design}}

With the goal of computing a QGNE of $\mathcal{G}^{\texttt{{sm}}}$
in mind, we capitalize on the potential structure of $\mathcal{G}^{\texttt{{sm}}}$
and construct the following multiplayer linearly constrained optimization
problem:\vspace{-0.1cm} 
\begin{equation}
(\mbox{P}):\,\begin{aligned} & \underset{\left(\vect p,{\vect p}^{\texttt{{J}}}\right)}{\text{maximize}}\quad r(\vect p,\vect p^{\texttt{{J}}})\triangleq{\displaystyle \sum}_{q=1}^{Q}\widetilde{{r_{q}^{s}}}(\mathbf{x}_{q})\\
 & \text{subject to}\quad\left(\vect p,{\vect p}^{\texttt{{J}}}\right)\in\mathcal{P}
\end{aligned}
\label{eq:SystemProblem-smooth}
\end{equation}
The above nonconcave maximization problem is smooth, thus the standard
definition of stationary solutions is applicable.

The connection between the social problem (P) in (\ref{eq:SystemProblem-smooth})
and the games $\mathcal{G}$ and $\mathcal{G}^{\texttt{{sm}}}$ is
given in the next proposition. \vspace{-0.1cm}

\begin{figure}[t]
\begin{centering}
\includegraphics[height=3.3cm]{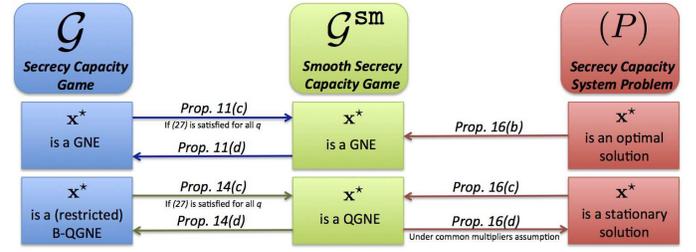} \\

\par\end{centering}

\caption{Connections between the games $\mathcal{G}$, $\mathcal{G}^{\texttt{{sm}}}$
and the social problem (P).}

\label{fig:connections} \vspace{-0.6cm}
\end{figure}

\begin{proposition} \label{prop:connections_social} Given $\mathcal{G}^{\text{\texttt{{sm}}}}$,
and the social problem (P) in (\ref{eq:SystemProblem-smooth}), the
following hold.

\item (a) If $\mathcal{P}\neq\emptyset$, then (P) has an optimal
solution;

\item (b) $[(P)\rightarrow\mathcal{G}^{\text{\texttt{{sm}}}}]$\emph{:}
If $\vect x^{\star}$ is an optimal solution of (P), then $\vect x^{\star}$
is a GNE of $\mathcal{G}^{\texttt{sm}}$;

\item (c) $[(P)\rightarrow\mathcal{G}^{\text{\texttt{{sm}}}}]$\emph{:}
If $\vect x^{\star}$ is a stationary solution of (P), then $\vect x^{\star}$
is a QGNE of $\mathcal{G}^{\texttt{sm}}$;

\item (d) $[\mathcal{G}^{\text{\texttt{{sm}}}}\rightarrow(P)]$\emph{:}
If $\vect x^{\star}$ is a QGNE of $\mathcal{G}^{\texttt{sm}}$ and
there exists common multipliers of the shared constraints $\sum_{r=1}^{Q}p_{jr}^{\texttt{{J}}}\leq P_{j}^{\texttt{{J}}}\quad j=1,\ldots,J$
for all players, then $\vect x^{\star}$ is a stationary solution
of (P). \end{proposition}\vspace{-0.2cm} 
\begin{IEEEproof}
See attached material. 
\end{IEEEproof}
Figure~\ref{fig:connections} summarizes the main results and relationship
between $\mathcal{G}$, $\mathcal{G}^{\texttt{{sm}}}$ and the social
problem (P), as stated in Propositions~\ref{prop:connections-GNE},
\ref{prop:connections_QGNE} and \ref{prop:connections_social}.

Based on Proposition \ref{prop:connections_social}, one can now design
distributed algorithms for computing a (Q)GNE of $\mathcal{G}^{\texttt{sm}}$
(and thus a restricted B-QGNE of the original game $\mathcal{G}$):
it is sufficient to solve the social problem (P). Such a problem is
nonconvex; stationary solutions however can be computed efficiently
observing that (P) is an instance of the DC program (\ref{eq:problem}),
under the following identifications:\vspace{-0.1cm} 
\[
\begin{aligned}f_{q}\left(p_{q},{\vect p}_{q}^{\texttt{{J}}}\right)\triangleq & -\log\left(\sigma^{2}+H_{qq}^{\texttt{SD}}p_{q}+{\textstyle \sum_{j=1}^{J}{H}_{jq}^{\texttt{JD}}\, p_{jq}^{\texttt{J}}}\right)\\
 & -\log\left(\sigma^{2}+{\textstyle \sum_{j=1}^{J}{H}_{je}^{\texttt{JE}}\, p_{jq}^{\texttt{J}}}\right)\\
g_{q}\left(p_{q},{\vect p}_{q}^{\texttt{{J}}}\right)\triangleq & -\log\left(\sigma^{2}+H_{qe}^{\texttt{SE}}p_{q}+{\textstyle \sum_{j=1}^{J}{H}_{je}^{\texttt{JE}}\, p_{jq}^{\texttt{J}}}\right)\\
 & -\log\left(\sigma^{2}+{\textstyle \sum_{j=1}^{J}{H}_{jq}^{\texttt{JD}}\, p_{jq}^{\texttt{J}}}\right).\vspace{-0.2cm}
\end{aligned}
\]

We can then efficiently compute a stationary solution of (P) using
any of the distributed algorithms introduced in Section \ref{sec:Distributed-Implementation}.
For instance, Algorithm \ref{alg:outer loop} based on a dual decomposition
loop (cf. Algorithm \ref{alg:Dual-Decomposition}) specialized to
(P) is given in Algorithm \ref{alg:SC}, where in (\ref{eq:users_update}),
$\mathcal{X}_{q}\triangleq\left\{ \vect x_{q}\triangleq(p_{q},\vect p_{q}^{\texttt{J}})\geq\vect0\,:\, p_{q}\leq P_{q}\text{ and (\ref{eq:linear_const})}\,\mbox{holds}\right\} $
and $\wt{\theta}_{q}(\vect x_{q};\vect x_{q}^{\nu})$ is defined as\vspace{-0.3cm}
\begin{equation}
\begin{aligned}\wt{\theta}_{q}(\vect x_{q};\vect x_{q}^{\nu})\triangleq f_{q}\left(\vect x_{q}\right)-\frac{\partial g_{q}(\vect x_{q}^{\nu})}{\partial p_{q}}\, p_{q}-\sum_{j=1}^{J}\frac{\partial g_{q}(\vect x_{q}^{\nu})}{\partial p_{jq}^{\texttt{J}}}\, p_{jq}^{\texttt{J}}\\
+\frac{\tau_{q}}{2}\|\vect x_{q}-\vect x_{q}^{\nu}\|^{2},
\end{aligned}
\vspace{-0.4cm}\label{eq:SC_theta_hat}
\end{equation}
with\vspace{-0.2cm} 
\[
\begin{aligned}\frac{\partial g_{q}(\vect x_{q})}{\partial p_{q}} & \triangleq-\frac{H_{qe}^{\texttt{SE}}}{\sigma^{2}+H_{qe}^{\texttt{SE}}p_{q}+{\sum_{j=1}^{J}{H}_{je}^{\texttt{JE}}\, p_{jq}^{\texttt{J}}}}\\
\frac{\partial g_{q}(\vect x_{q})}{\partial p_{jq}^{\texttt{J}}} & \triangleq-\frac{{H}_{je}^{\texttt{JE}}}{\sigma^{2}+H_{qe}^{\texttt{SE}}p_{q}+{\sum_{j=1}^{J}{H}_{je}^{\texttt{JE}}\, p_{jq}^{\texttt{J}}}}\\
 & \,\,\,\,\,\,\,-\frac{{H}_{jq}^{\texttt{JD}}}{\sigma^{2}+{\sum_{j=1}^{J}{H}_{jq}^{\texttt{JD}}\, p_{jq}^{\texttt{J}}}},\quad j=1,\ldots,J.
\end{aligned}
\]
\vspace{-0.3cm}\textcolor{red}{{} } 
\begin{algorithm}[h]
\textbf{Data}: $\boldsymbol{{\tau}}\triangleq(\tau_{q})_{q=1}^{Q}\geq\mathbf{0},\{\gamma^{\nu}\}>0$,
$\{\alpha^{t}\}>0$ and $\vect x^{0}\triangleq\left(\vect p^{0},{\vect p}^{\texttt{{J}},0}\right)\in\mathcal{P}$.
Set $\nu=0$.

$(\texttt{S.1})$: If $\vect x^{\nu}\triangleq\left(\vect p^{\nu},{\vect p}^{\texttt{{J}},\nu}\right)$
satisfies a termination criterion, STOP;

$(\texttt{S.2a})$: Choose $\boldsymbol{\lambda}^{0}\triangleq(\lambda_{j}^{0})_{j=1}^{J}\geq\vect0$.
Set $n=0$.

$(\texttt{S.2b})$: If $\boldsymbol{\lambda}^{t}\triangleq(\lambda_{j}^{t})_{j=1}^{J}$
satisfies a termination criterion, set $\wh{\vect x}\left(\vect x^{\nu}\right)=(p_{q}^{t,\,\nu},\vect p_{q}^{\texttt{J},\,\nu,\, t})_{q=1}^{Q}$
and go to $(\texttt{S.3})$;

$(\texttt{S.2c})$: The legitimate users $q=1,\ldots,Q$ compute in
parallel $(p_{q}^{\,\nu,\, t},\vect p_{q}^{\texttt{J},\,\nu,\, t})$
given by 
\begin{equation}
(p_{q}^{\,\nu,\, t},\vect p_{q}^{\texttt{J},\,\nu,\, t})\triangleq\underset{\vect x_{q}\triangleq(p_{q},\vect p_{q}^{\texttt{J}})\in\mathcal{X}_{q}}{\text{argmin}}\,\left\{ \wt{\theta}_{q}(\vect x_{q};\vect x_{q}^{\nu})+\boldsymbol{\lambda}^{t}{}^{T}\vect p_{q}^{\texttt{J}}\right\} ;\label{eq:users_update}
\end{equation}

$(\texttt{S.2d})$: Update $\boldsymbol{\lambda}\triangleq(\lambda_{j})_{j=1}^{J}$:
for all $j=1,\ldots,J$, 
\begin{equation}
\lambda_{j}^{t+1}\triangleq\left[0,\lambda_{j}^{t}+\alpha^{t}\left(\sum_{q=1}^{Q}p_{jq}^{\texttt{J},\,\nu,\, t}-P_{j}^{\texttt{J}}\right)\right]_{+};\label{eq:lambda_update}
\end{equation}

$(\texttt{S.2e})$: Set $t\leftarrow t+1$ and go back to $(\texttt{S.2b})$;

$(\texttt{S.3})$: Set $\vect x^{\nu+1}=\vect x^{\nu}+\gamma^{\nu}\left(\,\wh{\vect x}\left(\vect x^{\nu}\right)-\vect x^{\nu}\,\right)$;

$(\texttt{S.4})$: $\nu\leftarrow\nu+1$ and go to $(\texttt{S.1})$.

\caption{\hspace{-3pt}\textbf{:} \label{alg:SC}DC-based Algorithm for (P)}
\end{algorithm}

\vspace{-0.1cm}

If the sequences $\{\gamma^{\nu}\}>0$ and $\{\alpha^{t}\}>0$ are
chosen according to one of the rules stated in Theorems \ref{prop:convergence}
and \ref{thm:dual_convergence}, respectively, Algorithm \ref{alg:SC}
converges to a stationary solution of the social problem (P) (in the
sense of Theorem \ref{prop:convergence}), and thus to a QGNE of $\mathcal{G}^{\texttt{{sm}}}$
{[}cf. Proposition \ref{prop:connections_social}c){]}, which is also
a \textcolor{black}{restricted B-QGNE of $\mathcal{G}$} {[}cf. Proposition
\ref{prop:connections_QGNE}d){]}.\vspace{-0.2cm}

\begin{remark}[On the implementation of Algorithm \ref{alg:SC}]\rm
Once the CSI is available at the users' sides, Algorithm \ref{alg:SC}
can be implemented in a distributed way, with limited signaling only
between the legitimate users and the friendly jammers (no communications
among the users or the eavesdropper is required). Indeed, in the inner
loop of the algorithm, given the current value of the price $\boldsymbol{\lambda}$,
all the users simultaneously update the power profiles $(p_{q},\vect p_{q}^{\texttt{J}})$
solving locally a strongly convex optimization problem. Then, they
communicate to the friendly jammers the amount of power they need
resulting from the optimization. Given the power requests from the
users, the jammers update in parallel and independently their price
$\lambda_{j}$ performing an inexpensive scalar projection {[}cf.
(\ref{eq:lambda_update}){]}, and then broadcast the new price value
to the legitimate users. We remark that the proposed scheme require
the same CSI and communication overhead than CJ approaches proposed
in the literature \textcolor{black}{(see, e.g., \cite{GameTheoreticalModel-2,GameTheoreticalModel-6})}.\vspace{-0.2cm}
\end{remark}
\begin{remark}[More general formulation]\rm For the sake of simplicity, in the previous sections,  we assumed uniform power allocation over the spectrum (still to optimize) from the users and the friendly jammers. We remark however that game $\mathcal{G}$ in (\ref{eq:SCproblem}) [game  $\mathcal{G}^{\text{\texttt{{sm}}}}$ in (\ref{SCproblem-smooth}) and problem (P) in  (\ref{eq:SystemProblem-smooth})] can be generalized to the case of nonuniform power allocations (over flat-fading channels) and the proposed algorithms extended accordingly. We omit more details because of space limitation.  
\end{remark}
\vspace{-0.2cm}

\subsubsection{Numerical Results\label{sec:numerical}}

In this subsection, we present some experiments validating our theoretical
findings. We compare our Algorithm~\ref{alg:SC} with centralized
and applicable decentralized schemes existed in the literature (adapted to our formulation).

\noindent \emph{System setup}. All the experiments are obtained in
the following setting, unless stated otherwise. All the users and
jammers have the same power budget, i.e. $P_{q}=P_{j}^{\texttt{J}}=P$,
and we set $\texttt{snr}=P/\sigma^{2}=10$dB. The position of the
users, jammers, and eavesdropper are randomly generated within a square
area; the channel gains $H_{qq}^{\texttt{SD}}$, $H_{qe}^{\texttt{SE}}$,
$H_{jq}^{\texttt{JD}}$ and $H_{je}^{\texttt{JE}}$ are Rayleigh distributed
with mean equal to one and normalized by the (square) distance between
the transmitter and the receiver; our results are collected only for
the channel realizations satisfying condition (\ref{eq:linear_const}).
When present, there are $J=\lfloor Q/2\rfloor$ jammers. Algorithm~\ref{alg:SC}
is initialized by choosing the zero power allocation, and it is terminated
when the absolute value of the difference of the System Secrecy Rate
(SSR) in two consecutive iterations becomes smaller than $1e$-$5$.
Similarly, the inner loop is terminated when the difference of the
norm of the prices in two consecutive rounds is less than $1e$-$2$.

\begin{figure}[t]
\begin{centering}
\center\includegraphics[width=3.7in,height=2in]{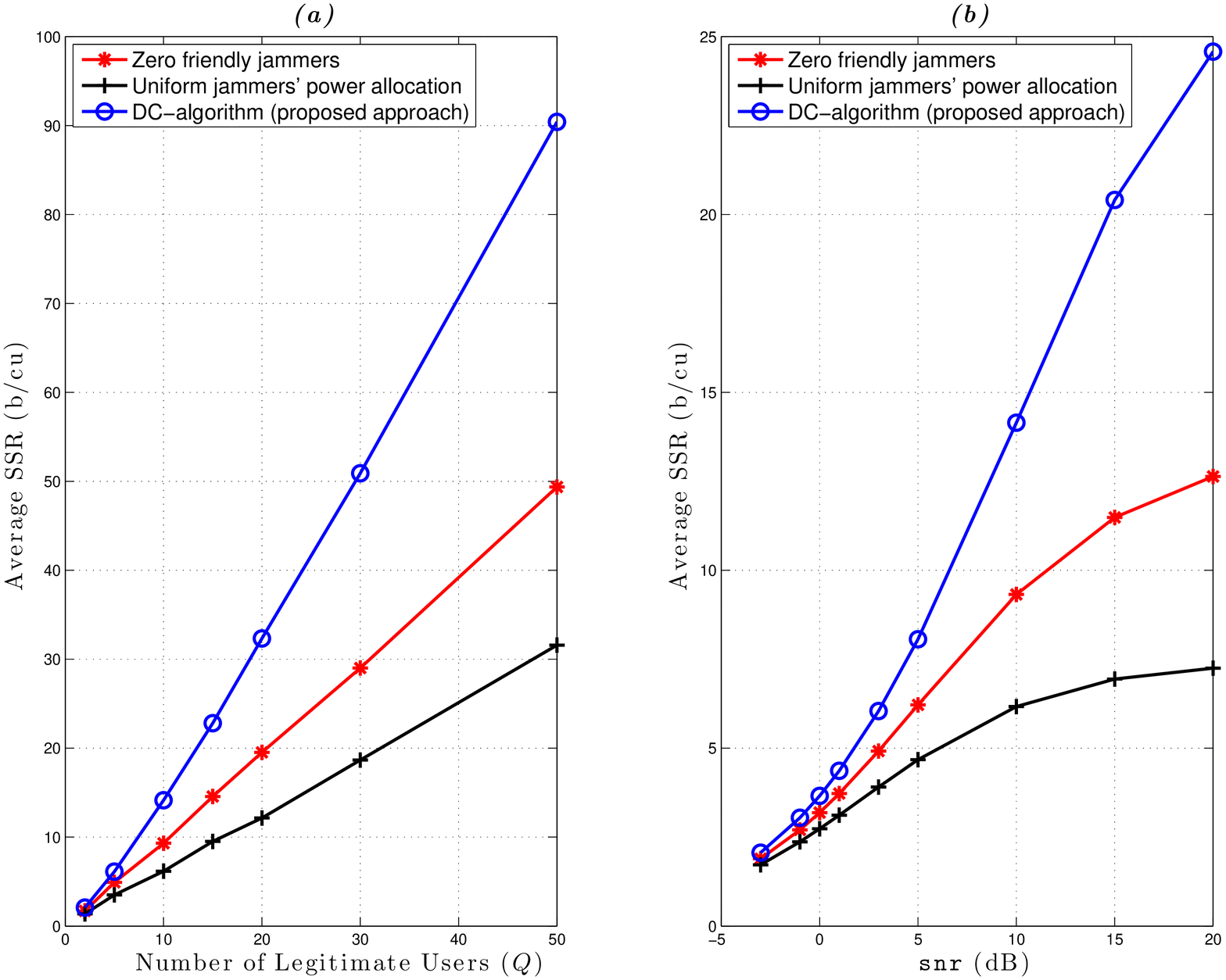}\vspace{-0.2cm}
 
\par\end{centering}

\caption{Average system secrecy rate (SSR) versus (a) number $Q$ of legitimate
users; (b) $\texttt{snr}$, for $Q=10$. 
}

\vspace{-0.5cm}
 \label{fig:SSRgain} 
\end{figure}

\emph{Example 1$-$Comparison with decentralized schemes}. In Fig.~\ref{fig:SSRgain}(a),
we plot the average SSR (taken over 50 independent channel realizations)
versus the number $Q$ of legitimate users achieved by our Algorithm~\ref{alg:SC}
(blue-line curves) and by solving the SSR maximization game while
assuming i) uniform power allocation for the jammers (black-line curves);
or ii) no friendly jammers available (red-line curve). In Fig.~\ref{fig:SSRgain}(b)
we plot the average SSR versus the $\texttt{snr}$ for the case of
$10$ main users; the rest of the setting is as in Fig.~\ref{fig:SSRgain}(a).
The figures show that the proposed approach yields much higher SSR
than that achievable by the other schemes, and the gain becomes more
significant as the number of users or the $\texttt{snr}$ increases.

\begin{figure}[t]
\begin{centering}
\includegraphics[width=3.7in,height=2in]{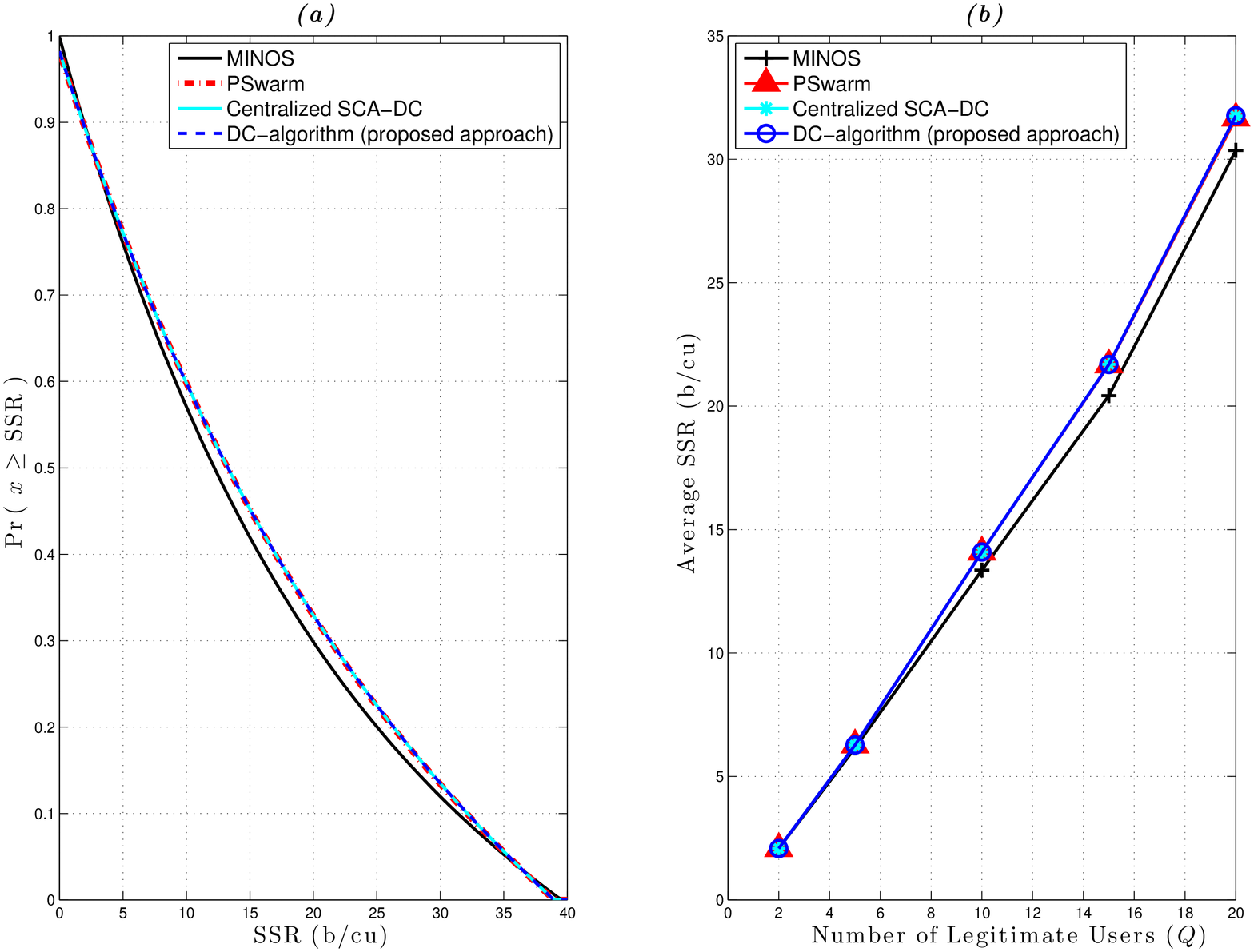}\vspace{-0.2cm}
 
\par\end{centering}

\caption{Comparison between our (distributed) Algorithm~\ref{alg:SC} and
centralized approaches.\label{fig:QualitySolution}\vspace{-0.7cm}
 }
\end{figure}

\emph{Example 2$-$Comparison with centralized schemes}. Since Algorithm~\ref{alg:SC}
converges to a stationary solution of the social problem (\ref{eq:SystemProblem-smooth}),
it is natural to compare our scheme with available centralized methods
attempting to compute locally or globally optimal solutions (but without
rigorously verifying their optimality). More specifically, we consider
two schemes: i) the NEOS server \cite{NEOSserver} based on MINOS
solver and PSwarm; and ii) the standard centralized SCA algorithm
for DC programs (see, e.g., \cite{DC-intro-3,SCA-3}). Even thought
these algorithms are computationally very demanding and not implementable
in a distributed network, they represent a good benchmark to test
our distributed algorithm. In Fig.~\ref{fig:QualitySolution}(a)
we plot the probability that the SSR exceeds a given value \texttt{SSR}
versus \texttt{SSR}, whereas in Fig.~\ref{fig:QualitySolution}(b)
we report the average SSR versus the number of legitimate users. All
the curves are computed running 100 independent experiments. The figures
show that our Algorithm~\ref{alg:SC} outperforms MINOS, and quite
surprisingly it has the same performance of PSwarm and SCA schemes.
This means that, at least for the experiments we simulated, Algorithm~\ref{alg:SC}
provides \emph{in a distributed way} solutions of (\ref{eq:SystemProblem-smooth})
that are very close to those obtained by centralized methods. 

\emph{Example 3$-$Convergence Speed}: Table~\ref{fig:Iterations}
shows the average number of inner and outer iterations required by
Algorithm~\ref{alg:SC} to converge versus the number of legitimate
users, when Rules 1 and 2 {[}c.f. (\ref{eq:step-size_diminishing})
and (\ref{eq:step-size_diminishing_2}){]} are used for the step-size
$\gamma^{\nu}$. The parameters in the two rules are set as $\beta_{1}=1e$-$6$,
$\beta_{2}=1e$-$4$, and $\epsilon=1e$-$5$. This average has been
taken over 50 independent channel realizations. Notice that, when
Rule 2 is used, the number of outer iterations is greatly reduced
in comparison to that obtained performing Rule 1. Note also that,
the average number of inner iterations per outer iteration is slightly
increased in the former case; \textcolor{black}{however, overall,
Algorithm~\ref{alg:SC} with Rule 2 seems to be faster than the same
algorithm based on Rule 1. As far as the quality of the achieved SSR
is concern, we observed results consistent to }Fig.~\ref{fig:QualitySolution}\textcolor{black}{.
\vspace{-0.1cm}
}\textcolor{black}{\vspace{-.1cm}}{\setlength{\tabcolsep}{2.8 pt}   \begin{table}[h] {\scriptsize } \begin{tabular}{|c|c|c|c|c|c|c|c|} \hline {{\sl \scriptsize number of  users ($Q$)}} & \multicolumn{1}{c|}{{\sl \scriptsize 2}} & \multicolumn{1}{c|}{{\sl \scriptsize 5}} & \multicolumn{1}{c|}{{\sl \scriptsize 10}} & \multicolumn{1}{c|}{{\sl \scriptsize 15}} & \multicolumn{1}{c|}{{\sl \scriptsize 20}}  & \multicolumn{1}{c|}{{\sl \scriptsize 30}} & \multicolumn{1}{c|}{{\sl \scriptsize 50}}\tabularnewline \hline  {\scriptsize \mbox{}\emph{\scriptsize Rule 1}{\scriptsize{} - outer iterations} } & {\scriptsize 46.80 } & {\scriptsize 53.62 } & {\scriptsize 73.52 } & {\scriptsize 95.48 }  & {\scriptsize 97.74 } & {\scriptsize 138.00 } & {\scriptsize 175.56}\tabularnewline {\scriptsize \mbox{}\emph{\scriptsize Rule 1}{\scriptsize{} - inner iterations } } & {\scriptsize 1.02 } & {\scriptsize 1.08 } & {\scriptsize 1.19 } & {\scriptsize 1.23 } & {\scriptsize 1.27 } & {\scriptsize 1.23 } & {\scriptsize 1.21 }\tabularnewline \hline  {\scriptsize \mbox{}\emph{\scriptsize Rule 2}{\scriptsize{} - outer iterations} } & {\scriptsize 46.66 } & {\scriptsize 56.58 } & {\scriptsize 66.78 } & {\scriptsize 76.12 } & {\scriptsize 72.42 } & {\scriptsize 83.44 } & {\scriptsize 89.56 }\tabularnewline {\scriptsize \mbox{}\emph{\scriptsize Rule 2}{\scriptsize{} - inner iterations } } & {\scriptsize 1.02 } & {\scriptsize 1.08 } & {\scriptsize 1.20 } & {\scriptsize 1.28 } & {\scriptsize 1.30 }  & {\scriptsize 1.30 } & {\scriptsize 1.33 }\tabularnewline \hline  \end{tabular}{\scriptsize \caption{{\scriptsize Average number of outer iterations and inner iterations per outer iteration in Algorithm~\ref{alg:SC} for different choices of diminishing step-size rules.\label{fig:Iterations}}}\vspace{-0.5cm} } \end{table}\

\subsection{Case Study \#2: Design of CR MIMO systems\label{CR_problem}}

\subsubsection{System model and problem formulation\label{sub:System-model_CR}}

Consider a hierarchical MIMO CR system composed of $P$ PUs sharing
the licensed spectrum with $I$ SUs; the network of SUs is modeled
as a $I$-user vector Gaussian interference channel. Each SU $i$
is equipped with $n_{T_{i}}$ and $n_{R_{i}}$ transmit and receive
antennas, respectively, and PUs may have multiple antennas. Let $\mathbf{H}_{ji}\!\in\!\mathbb{C}^{n_{R_{j}}\times n_{T_{i}}}$
(resp. $\mathbf{G}_{pi}$) be the cross-channel matrix between the
secondary transmitter $i$ and the secondary receiver $j$ (resp.
primary receiver $p$). Under basic information theoretical assumptions,
the transmission rate of SU $i$ is
\begin{equation}
r_{i}(\mathbf{Q}_{i},\mathbf{Q}_{-i})\triangleq\log\det\left(\mathbf{I}+\mathbf{H}_{ii}^{H}\mathbf{R}_{i}(\mathbf{Q}_{-i})^{-1}\mathbf{H}_{ii}\mathbf{Q}_{i}\right)\label{rmg:rt}
\end{equation}
where $\mathbf{Q}_{i}$ is the transmit covariance matrix of SU $i$
(to be optimized), $\mathbf{Q}_{-i}\triangleq(\mathbf{Q}_{j})_{j\neq i}$,
$\mathbf{R}_{i}(\mathbf{Q}_{-i})\triangleq\mathbf{R}_{n_{i}}+\sum_{j\neq i}\mathbf{H}_{ij}\mathbf{Q}_{j}\mathbf{H}_{ij}^{H}$
with $\mathbf{R}_{n_{i}}\succ\mathbf{0}$ being the covariance matrix
of the noise plus the interference generated by the active PUs. Each
SU $i$ is subject to the following local constraints 
\begin{equation}
\mathcal{Q}_{i}\triangleq\left\{ \mathbf{Q}_{i}\succeq\mathbf{0}:\begin{array}{l}
\mathrm{tr}(\mathbf{Q}_{i})\leq P_{i}^{\mathrm{tot}},\;\mathbf{Q}_{i}\in\mathcal{Z}_{i}\end{array}\right\} ,\label{rmg:pc}
\end{equation}
where $P_{i}^{\mathrm{tot}}$ is the total transmit power and $\mathcal{Z}_{i}\subseteq\mathbb{C}^{n_{i}\times n_{i}}$
is an abstract closed and convex set suitable to accommodate (possibly)
additional local constraints, such as: i) \emph{null constraints}
$\mathbf{U}_{i}^{H}\mathbf{Q}_{i}=\mathbf{0}$, with $\mathbf{U}_{i}$
being a $n_{T_{i}}\times r_{i}$ (with $r_{i}<n_{T_{i}}$), which
prevents SUs to transmit along some prescribed ``directions'' (the
columns of $\mathbf{U}_{i}$); and ii)\emph{ soft and peak power shaping
constraints} in the form of $\mathrm{tr}(\mathbf{T}_{i}^{H}\mathbf{Q}_{i}\mathbf{T}_{i})\leq I_{i}^{\text{{ave}}}$
and $\lambda^{\max}(\mathbf{F}_{i}^{H}\mathbf{Q}_{i}\mathbf{F}_{i})\leq I_{i}^{\text{{peak}}}$,
which limits to $I_{i}^{\text{{ave}}}>0$ and $I_{i}^{\text{{peak}}}>0$
the total average and peak average power allowed to be radiated along
the range space of matrices $\mathbf{T}_{i}\in\mathbb{C}^{n_{T_{i}}\times n_{G_{i}}}$
and $\mathbf{F}_{i}\in\mathbb{C}^{n_{T_{i}}\times n_{F_{i}}}$, respectively.
Adopting the spectrum underlay architecture, the interference power
at the primary receivers is regulated by imposing global interference
constraints to the SUs, in the form of $\sum_{i}\mathrm{tr}(\mathbf{G}_{pi}^{H}\mathbf{Q}_{i}\mathbf{G}_{pi})\leq I_{p}^{\text{{tot}}}$
for all $p=1,\ldots,P,$ where $I_{p}^{\text{{tot}}}>0$ is the interference
threshold imposed by PU $p$.

The design of the CR system can be formulated as 
\begin{equation}
\!\!\!\!\begin{array}{cl}
\underset{\mathbf{Q}_{1},\ldots,\mathbf{Q}_{I}}{\textrm{maximize}} & {\theta}(\mathbf{Q})\triangleq\sum_{i=1}^{I}r_{i}(\mathbf{Q}_{i},\mathbf{Q}_{-i})\\
\textrm{subject to} & \mathbf{Q}_{i}\in\mathcal{Q}_{i},\quad i=1,\ldots,I,\smallskip\\
 & \sum_{i}\mathrm{tr}(\mathbf{G}_{pi}^{H}\mathbf{Q}_{i}\mathbf{G}_{pi})\leq I_{p}^{\text{{tot}}},\quad p=1,\ldots,P.
\end{array}\label{eq:sum-rate-1}
\end{equation}

Special cases of the nonconvex sum-rate maximization problem (\ref{eq:sum-rate-1})
have already been studied in the literature, e.g., in \cite{Kim2011,DallAnese2012}.
However the theoretical convergence of current algorithms is up to
date an open problem. Since (\ref{eq:sum-rate-1}) is an instance
of (\ref{eq:problem}), we can capitalize on the framework developed
in the first part of the paper and obtain readily a class of distributed
algorithms with \emph{provable convergence}.

\subsubsection{DC-based decomposition algorithms}

We cast first (\ref{eq:sum-rate-1}) into (\ref{eq:problem}) (in
the maximization form). Exploring the DC-structure of the rates $r_{i}(\mathbf{Q}_{i},\mathbf{Q}_{-i})$,
the sum-rate ${\theta}(\mathbf{Q})$ can indeed be rewritten as the
sum of a concave and convex function, namely: ${\theta}(\mathbf{Q})=\sum_{i}\left(f_{i}(\mathbf{Q})-g_{i}(\mathbf{Q})\right)$,
where 
\begin{eqnarray*}
f_{i}(\mathbf{Q}) & \!\!\!\!\triangleq\,\,\,\log\det\left(\mathbf{R}_{n_{i}}+\sum_{j=1}\mathbf{H}_{ij}\mathbf{Q}_{j}\mathbf{H}_{ij}^{H}\right)\\
g_{i}(\mathbf{Q}_{-i}) & \!\!\!\!\triangleq\,\,\log\det\left(\mathbf{R}_{n_{i}}+\sum_{j\neq i}\mathbf{H}_{ij}\mathbf{Q}_{j}\mathbf{H}_{ij}^{H}\right).
\end{eqnarray*}
We can now use the machinery developed in Sec. \ref{sub:SCA-centralized}.
It is not difficult to show that the approximation function $\widetilde{{\theta}}(\vect x;\vect x^{\nu})$
in (\ref{theta hat1}) (to be maximized) becomes (up to a constant
term) $\widetilde{{\theta}}(\mathbf{Q};\mathbf{Q}^{\nu})\triangleq\sum_{i}\widetilde{{\theta}}_{i}(\mathbf{Q};\mathbf{Q}^{\nu})$,
with 
\begin{equation}
\!\!\widetilde{{\theta}}_{i}(\mathbf{Q};\mathbf{Q}^{\nu})\!\triangleq \! r_{i}(\mathbf{Q}_{i},\mathbf{Q}_{-i}^{\nu})-\left\langle \mathbf{\Pi}_{i}(\mathbf{Q}^{\nu}),\mathbf{Q}_{i}\!-\!\mathbf{Q}_{i}^{\nu}\right\rangle -\tau_{i}\left\Vert \mathbf{Q}_{i}-\mathbf{Q}_{i}^{\nu}\right\Vert _{F}^{2}\label{eq:teta_tilde_MIMO}
\end{equation}
where $\mathbf{Q}^{\nu}\triangleq(\mathbf{Q}_{i}^{\nu})_{i=1}^{I}$
with each $\mathbf{Q}_{i}^{\nu}\succeq\mathbf{0}$, $\left\langle \mathbf{A},\mathbf{B}\right\rangle \triangleq\text{{Re}}\left\{ \text{{tr}}(\mathbf{A}^{H}\mathbf{B})\right\} $,
$\left\Vert \bullet\right\Vert _{F}$ is the Frobenius norm, and 
\begin{equation}
\mathbf{\Pi}_{i}(\mathbf{Q}^{\nu})\triangleq\sum_{j\in\mathcal{N}_{i}}\,\mathbf{H}_{ji}^{H}\,\widetilde{\mathbf{R}}_{j}(\mathbf{Q}_{-j}^{\nu})\,\mathbf{H}_{ji},\vspace{-0.1cm}\label{eq:pricing_term}
\end{equation}
with $\mathcal{N}_{i}$ denoting the set of neighbors of user $i$
(i.e., the set of users $j$'s which user $i$ interferers with),
and 
\[
\widetilde{\mathbf{R}}_{j}(\mathbf{Q}_{-j}^{\nu})\triangleq\mathbf{R}_{j}(\mathbf{Q}_{-j}^{\nu})^{-1}-(\mathbf{R}_{j}(\mathbf{Q}_{-j}^{\nu})+\mathbf{H}_{jj}\mathbf{Q}_{j}^{\nu}\mathbf{H}_{jj}^{H})^{-1}.
\]

Therefore a stationary solution of (\ref{eq:sum-rate-1}) can be efficiently
computed using any of the distributed algorithms introduced in Section
\ref{sec:Distributed-Implementation}; one just needs to replace $\widetilde{{\theta}}(\vect x;\vect x^{\nu})$
with $\widetilde{{\theta}}(\mathbf{Q};\mathbf{Q}^{\nu})$ (and the
minimization with the maximization). For instance, Algorithm \ref{alg:outer loop}
based on a dual decomposition loop (cf. Algorithm \ref{alg:Dual-Decomposition})
can be written in the form of Algorithm \ref{alg:Dual-Decomposition_MIMO}.
Convergence (in the sense of Theorem \ref{prop:convergence}) is guaranteed
if the step-size sequences $\{\gamma^{\nu}\}$ and $\{\alpha^{n}\}>0$
are chosen according to one of the rules stated in Theorems \ref{prop:convergence}
and \ref{thm:dual_convergence}, respectively. \vspace{-0.1cm} 
\begin{algorithm}[H]
\textbf{Data}: $\boldsymbol{{\tau}}\triangleq(\tau_{i})_{i=1}^{I}\geq\mathbf{0},\{\gamma^{\nu}\}>0$,
$\{\alpha^{t}\}>0$ and $\mathbf{Q}_{i}^{0}\in\mathcal{Q}$ for all
$i$. Set $\nu=0$.

$(\texttt{S.1})$: If $\mathbf{Q}^{\nu}\!\!\triangleq\!\!(\mathbf{Q}_{i}^{\nu})_{i=1}^{I}$
satisfies a termination criterion, STOP;

$(\texttt{S.2a})$: Choose $\boldsymbol{\lambda}^{0}\triangleq(\lambda_{p}^{0})_{p=1}^{P}\geq\vect0$.
Set $t=0$.

$(\texttt{S.2b})$: If $\boldsymbol{\lambda}^{t}\triangleq(\lambda_{p}^{t})_{p=1}^{P}$
satisfies a termination criterion, set $\wh{\vect Q}\left(\vect Q^{\nu}\right)\triangleq(\mathbf{Q}_{i}^{t,\,\nu})_{i=1}^{I}$
and go to $(\texttt{S.3})$;

$(\texttt{S.2c})$: The SUs solve in parallel the following strongly
convex optimization problems: for all $i=1,\dots,I$,\vspace{-0.1cm}
\begin{equation}
\mathbf{Q}_{i}^{\,\nu,\, t}\triangleq\underset{\vect Q_{i}\in\mathcal{Q}_{i}}{\text{argmax}}\left\{ \wt{\theta}_{i}(\mathbf{Q}_{i};\mathbf{Q}^{\nu})-\sum_{p=1}^{P}\lambda_{p}^{t}\,\mathrm{tr}(\mathbf{G}_{pi}^{H}\mathbf{Q}_{i}\mathbf{G}_{pi})\right\} ;\label{eq:users_update-1}
\end{equation}

$(\texttt{S.2d})$: Update $\boldsymbol{\lambda}\triangleq(\lambda_{p})_{p=1}^{P}$:
for all $p=1,\ldots,P$, 
\begin{equation}
\lambda_{p}^{t+1}\triangleq\left[\lambda_{p}^{t}+\alpha^{t}\left(\sum_{i=1}^{I}\mathrm{tr}(\mathbf{G}_{pi}^{H}\mathbf{Q}_{i}^{\,\nu,\, t}\mathbf{G}_{pi})-I_{p}^{\max}\right)\right]_{+};\label{eq:lambda_update-1}
\end{equation}

$(\texttt{S.2e})$: Set $t\leftarrow t+1$ and go back to $(\texttt{S.2b})$;

$(\texttt{S.3})$: Set $\vect Q^{\nu+1}=\vect Q^{\nu}+\gamma^{\nu}\left(\,\wh{\vect Q}\left(\vect Q^{\nu}\right)-\vect Q^{\nu}\,\right)$;

$(\texttt{S.4})$: $\nu\leftarrow\nu+1$ and go to $(\texttt{S.1})$.

\caption{\hspace{-3pt}\textbf{: \label{alg:Dual-Decomposition_MIMO}}Dual-based
Algorithm for ({\small \ref{eq:sum-rate-1}}).}
\end{algorithm}
\vspace{-0.2cm}

\subsubsection{Discussion on the implementation of Algorithm \ref{alg:Dual-Decomposition_MIMO} }

The algorithm is a double loop scheme in the sense described next. 

\noindent\emph{ Inner loop}: In this loop, at every iteration $t$:
i) First, all SUs solve in \emph{parallel} their strongly convex optimization
problems (\ref{eq:users_update-1}), for fixed $\boldsymbol{{\lambda}}^{t}=(\lambda_{p}^{t})_{p=1}^{P}$,
resulting in the optimal solutions $(\mathbf{Q}_{i}^{\,\nu,\, t})_{i=1}^{I}$;
ii) Then, given the new interference levels $\sum_{i=1}^{I}\mathrm{tr}(\mathbf{G}_{pi}^{H}\mathbf{Q}_{i}^{\,\nu,\, t}\mathbf{G}_{pi})$,
the prices $\boldsymbol{{\lambda}}$ are updated in parallel via (\ref{eq:lambda_update-1}),
resulting in $\boldsymbol{{\lambda}}^{t+1}=(\lambda_{p}^{t+1})_{p=1}^{P}$.
The loop terminates when $\{\boldsymbol{\lambda}^{t}\}$ meets the
termination criterion in (S.2b). 

\noindent\emph{ Outer loop}: It consists in updating $\mathbf{Q}_{i}^{\nu}$'s
according to (S.3).

\noindent\emph{ Communication overhead}: The proposed algorithm is
fairly distributed. Indeed, given the interference generated by the
other users {[}the covariance matrix $\mathbf{R}_{i}(\mathbf{Q}_{-i})$,
which can be locally measured{]}, and the interference price $\boldsymbol{{\Pi}}_{i}(\mathbf{Q}^{\nu})$,
each SU can efficiently and \emph{locally} compute the optimal covariance
matrix $\mathbf{Q}_{i}^{\,\nu,\, t}$ by solving (\ref{eq:users_update-1}).
Note that, for some specific structures of the feasible sets $\mathcal{Q}_{i}$
and channels (e.g., $\mathcal{Z}_{i}=\emptyset$, full-column rank
channel matrices $\mathbf{H}_{ii}$, and $\tau_{i}=0$), a solution
of (\ref{eq:users_update-1}) is available in closed form (up to the
multipliers associated with the power budget constraints) \cite{Kim2011}.
The estimation of the prices $\boldsymbol{{\Pi}}_{i}(\mathbf{Q}^{\nu})$
requires some signaling exchange but only among nearby users. Interestingly,
the pricing expression (\ref{eq:pricing_term}) as well as the signaling
overhead necessary to compute it coincides with that of pricing schemes
proposed in the literature to solve related problems \cite{Kim2011,DDPA-2,SchmidtShiBerryHonigUtschick-SPMag}. 

The natural candidates for updating the prices in the inner loop are
the PUs, after measuring \emph{locally} the current overall interference
$\sum_{i=1}^{I}\mathrm{tr}(\mathbf{G}_{pi}^{H}\mathbf{Q}_{i}^{\,\nu,\, t}\mathbf{G}_{pi})$
from the SUs. Note that this update is computationally inexpensive (it
is a projection onto $\mathbb{R}_{+}$), can be performed in parallel
among PUs, and does not require any signaling exchange with the SUs.
The new value of the prices is then broadcasted to the SUs. In CR
scenarios where the PUs cannot participate in the updating process,
the SUs themselves can perform the price update, at the cost of more
signaling, e.g., using consensus algorithms. Alternatively, if the
primary receivers have a fixed geographical location, it might be
possible to install some monitoring devices close to each primary
receiver having the functionality of price computation and broadcasting. 

As a final remark note that, since Algorithm \ref{alg:Dual-Decomposition_MIMO}
is a dual-base scheme, it is scalable with the number of SUs. 
However, for the same reasons, there might happen that the interference
constraints are not satisfied during the intermediate iterations.
This issue can be alleviated in practice by choosing a ``large''
$\boldsymbol{{\lambda}}^{0}$ as initial price. An alternative distributed
scheme which does not suffer from this issue can be readily obtained
using the primal-based decomposition approach introduced in Sec. \ref{sub:Primal-decomposition};
we omit the details because of space limitation.

\section{Conclusions\label{sec:conclusion}}

In this paper we proposed a novel decomposition framework to compute
stationary solutions of nonconvex (possibly DC) sum-utility minimization
problems with coupling convex constraints. We developed a class of
(inexact) best-response-like algorithms, where all the users iteratively
solve in \emph{parallel} a suitably convexified version of the original
DC program. To the best of our knowledge, this is the first set of
\emph{distributed }algorithms \emph{with provable convergence} for
multiuser DC programs with \emph{coupling constraints}. Finally, we
tested our methodology on two problems: i) a novel secrecy rate game
(for which we provided a nontrivial DC formulation); and ii) the sum-rate
maximization problems over CR MIMO networks. Our experiments show
that our distributed algorithms reach performance comparable (and
sometimes better) than centralized schemes.

\appendices{}

\section{\label{sec:Appendix:-Theorem_convergence}Proof of Theorem \ref{prop:convergence}}

The proof follows from \cite[Th.3]{DDPA-2} and Proposition~\ref{prop:map properties}
below, which establishes the main properties of the best-response
map $\Xi\ni\vect y\mapsto\wh{\vect x}(\vect y)\in\Xi$ defined in
(\ref{eq:map}), as required by \cite[Th.3]{DDPA-2}.

\begin{proposition}\label{prop:map properties} Given the DC program
(\ref{eq:problem}) under A1-A4, the map $\Xi\ni\vect y\mapsto\wh{\vect x}(\vect y)\in\Xi$
defined in (\ref{eq:map}) has the following properties:

\noindent \emph{(a):} For every given $\vect y\in\Xi$, the vector
$\wh{\vect x}(\vect y)-\vect y$ is a descent direction of the objective
function $\theta(\vect x)$ at $\vect y$: 
\begin{equation}
\left(\wh{\vect x}(\vect y)-\vect y\right)^{T}\nabla_{\vect x}\theta(\vect y)\leq-c_{\boldsymbol{{\tau}}}\,\|\wh{\vect x}(\vect y)-\vect y\|^{2},\label{eq:descent_direction}
\end{equation}
with $c_{\boldsymbol{\tau}}$ defined in (\ref{eq:descent_constant-1});
\smallskip{}

\noindent \emph{(b):} $\wh{\vect x}(\bullet)$ is Lipschitz continuos
on $\Xi$, with constant \textcolor{black}{$L_{\wh{\vect x}}\triangleq L_{\nabla\widetilde{\theta}}/c_{\boldsymbol{\tau}}$},
where $L_{\nabla\widetilde{\theta}}$ is defined in Lemma \ref{lem:lipschitz_par}
below; \smallskip{}

\noindent \emph{(c):} The set of fixed points of $\wh{\vect x}(\bullet)$
coincides with the set of stationary points of the optimization problem
(\ref{eq:problem}); therefore $\wh{\vect x}(\bullet)$ has a fixed
point.\end{proposition}

\emph{Proof.} Before proving the proposition, we introduce the following
lemma, whose proof is omitted (see attached material).

\begin{lemma}\label{lem:lipschitz_par}In the setting of Proposition
\ref{prop:map properties}, $\nabla\wt{\theta}({\vect x};\bullet)$
is uniformly Lipschitz on $\Xi$, that is, for any given $\mathbf{x}\in\Xi$,
\begin{equation}
\left\Vert \nabla_{\vect x}\wt{\theta}({\vect x};\vect y)-\nabla_{\vect x}\wt{\theta}({\vect x};\vect z)\right\Vert \leq L_{\nabla\wt{\theta}}\,\left\Vert \vect z-\vect y\right\Vert ,\quad\forall\vect y,\vect z\in\Xi,\label{eq:Lemma_row1}
\end{equation}
with \textcolor{black}{$L_{\nabla\wt{\theta}}^{2}\triangleq4\,(L_{\nabla\theta}^{2}+2\sum_{i=1}^{I}L_{\nabla f_{i}}^{2}+\tau^{\max})$},
where $L_{\nabla\theta}$ and $L_{\nabla f_{i}}$ are defined in assumption
A3, and $\tau^{\max}\triangleq\max_{i}\tau_{i}^2$.\end{lemma}

We prove next only (a) and (b) of Proposition~\ref{prop:map properties}.\emph{ }

\emph{(a)}: Given $\vect y\in\Xi$, by definition, $\wh{\vect x}(\vect y)$
satisfies the minimum principle associated with (\ref{eq:map}): for
all $\vect z\triangleq\left(\vect z_{i}\right)_{i=1}^{I}\in\Xi$,\vspace{-0.1cm}
 
\begin{equation}
\begin{aligned}\left(\vect z-\wh{\vect x}(\vect y)\right)^{T}\nabla_{\vect x}\wt{\theta}(\wh{\vect x}(\vect y);\vect y)\geq0\\
\sum_{i=1}^{I}\left(\vect z_{i}-\wh{\vect x}_{i}(\vect y)\right)^{T}\left[\nabla_{\vect x_{i}}f_{i}(\wh{\vect x}_{i}(\vect y),\vect y_{-i})+\sum_{j\neq i}\nabla_{\vect x_{i}}f_{j}(\vect y)\right.\\
\left.-\sum_{j=1}^{I}\nabla_{\vect x_{i}}g_{j}(\vect y)+\tau_{i}\left(\wh{\vect x}_{i}(\vect y)-\vect y_{i}\right)\right]\geq0.
\end{aligned}
\label{eq:min_princ_teta_tilde}
\end{equation}
Letting $\vect z_{i}=\vect y_{i}$, and, adding and subtracting $\nabla_{\vect x_{i}}f_{i}(\vect y)$
in each term $i$ of the sum in (\ref{eq:min_princ_teta_tilde}),
we obtain: 
\begin{equation}
\hspace{-0.1cm}\begin{aligned} & \left(\vect y-\wh{\vect x}(\vect y)\right)^{T}\nabla_{\vect x}\theta(\vect y)\\
 & \geq\sum_{i=1}^{I}\left(\wh{\vect x}_{i}(\vect y)-\vect y_{i}\right)^{T}\left(\nabla_{\vect x_{i}}f_{i}(\wh{\vect x}_{i}(\vect y),\vect y_{-i})-\nabla_{\vect x_{i}}f_{i}(\vect y)\right)\\
 & \,\,\,\,\,\,\,+\sum_{i=1}^{N}\tau_{i}\|\wh{\vect x}_{i}(\vect y)-\vect y_{i}\|^{2}\geq c_{\boldsymbol{{\tau}}}\,\|\wh{\vect x}(\vect y)-\vect y\|^{2},\vspace{0.2cm}
\end{aligned}
\label{eq:descent_direction_proof}
\end{equation}
where in the last inequality we used the definition of $c_{\boldsymbol{\tau}}$. This completes the proof
of (a).

\noindent \emph{(b)}: Let $\vect y,\vect z\in\Xi$; by the minimum
principle, we have 
\begin{eqnarray}
\left(\vect{v}-\wh{\vect{x}}(\vect{y})\right)^{T}\nabla_{{\vect{x}}}\wt{\theta}(\wh{\vect{x}}(\vect{y});\vect{y}) & \geq & 0\qquad\forall\vect{v}\in\Xi\nonumber \\
\left(\vect{w}-\wh{\vect{x}}(\vect{z})\right)^{T}\nabla_{{\vect{x}}}\wt{\theta}(\wh{\vect{x}}(\vect{z});\vect{z}) & \geq & 0\qquad\forall\vect{w}\in\Xi.\label{eq:mp2}
\end{eqnarray}
Setting $\vect v=\wh{\vect x}(\vect z)$ and $\vect w=\wh{\vect x}(\vect y)$
and summing the two inequalities above, after some manipulations,
we obtain:\vspace{-0.1cm} 
\begin{equation}
\begin{array}{l}
\left(\wh{\vect{x}}(\vect{z})-\wh{\vect{x}}(\vect{y})\right)^{T}\left(\nabla_{{\vect{x}}}\wt{\theta}(\wh{\vect{x}}(\vect{z});\vect{z})-\nabla_{{\vect{x}}}\wt{\theta}(\wh{\vect{x}}(\vect{y});\vect{z})\right)\\
\leq\left(\wh{\vect{x}}(\vect{y})-\wh{\vect{x}}(\vect{z})\right)^{{T}}\left(\nabla_{{\vect{x}}}\wt{\theta}(\wh{\vect{x}}(\vect{y});\vect{z})-\nabla_{{\vect{x}}}\wt{\theta}(\wh{\vect{x}}(\vect{y});\vect{y})\right).
\end{array}\label{eq:minimum_principle_Lip}
\end{equation}
The Lipschitz property of $\wh{\vect x}(\bullet)$, as stated in Proposition
\ref{prop:map properties}(b), comes from (\ref{eq:minimum_principle_Lip})
using the following lower and upper bounds: 
\begin{equation}
\begin{array}{l}
\left(\wh{\vect{x}}(\vect{z})-\wh{\vect{x}}(\vect{y})\right)^{{T}}\left(\nabla_{{\vect{x}}}\wt{\theta}(\wh{\vect{x}}(\vect{z});\vect{z})-\nabla_{{\vect{x}}}\wt{\theta}(\wh{\vect{x}}(\vect{y});\vect{y})\right)\\
\geq c_{{\boldsymbol{\tau}}}\left\Vert \wh{\vect{x}}(\vect{z})-\wh{\vect{x}}(\vect{y})\right\Vert ^{{2}}\vspace{-0.1cm}
\end{array}\label{eq:lipschtz_map_2}
\end{equation}
and\vspace{-0.1cm} 
\begin{equation}
\begin{array}{l}
\left(\wh{\vect{x}}(\vect{y})-\wh{\vect{x}}(\vect{z})\right)^{{T}}\left(\nabla_{{\vect{x}}}\wt{\theta}(\wh{\vect{x}}(\vect{y});\vect{z})-\nabla_{{\vect{x}}}\wt{\theta}(\wh{\vect{x}}(\vect{y});\vect{y})\right)\\
\leq\, L_{\nabla\wt{\theta}}\,\left\Vert \wh{\vect{x}}(\vect{y})-\wh{\vect{x}}(\vect{z})\right\Vert \,\left\Vert \vect{z}-\vect{y}\right\Vert 
\end{array}\label{eq:lipschtz_map_1}
\end{equation}
where (\ref{eq:lipschtz_map_2}) is a direct consequence of the uniform
strong convexity of $\wt{\theta}$, whereas (\ref{eq:lipschtz_map_1})
follows from the Cauchy-Schwartz inequality and Lemma \ref{lem:lipschitz_par}.
Combining (\ref{eq:minimum_principle_Lip}), (\ref{eq:lipschtz_map_2}),
and (\ref{eq:lipschtz_map_1}), we obtain the desired result.

\vspace{-0.3cm}

\section{Proof of Proposition \ref{prop:connections-GNE}\label{Proof-of-Proposition_GNE}}

\noindent (a) It is not difficult to check that $\mathcal{G}$ is
an exact potential game with potential function $\Phi(\mathbf{p},\mathbf{p}^{\texttt{{J}}})\triangleq\sum_{q}r_{q}^{s}(p_{q},\mathbf{p}_{q}^{\texttt{{J}}})$.
It turns out that any optimal solution of the associated multiplayer
maximization problem: 
\begin{equation}
\begin{aligned}\begin{array}{ll}
\underset{\left(\vect p,{\vect p}^{\texttt{{J}}}\right)\geq\mathbf{0}}{\text{maximize}} & \Phi(\mathbf{p},\mathbf{p}^{\texttt{{J}}})\\
\text{subject to:} & \begin{array}[t]{c}
\vspace{-0.4cm}\\
\hspace{-0.2cm}\hspace{-0.2cm}\hspace{-0.2cm}\begin{array}[t]{l}
\begin{array}{ll}
p_{q}\leq P_{q}, & \forall q=1,\ldots,Q\\
{\displaystyle {\sum_{r=1}^{Q}}}\: p_{jr}^{\texttt{{J}}}\leq P_{j}^{\texttt{{J}}}, & \forall\, j=1,\ldots,J,
\end{array}\end{array}
\end{array}
\end{array}\end{aligned}
\label{eq:GNEexistence}
\end{equation}
is a GNE of $\mathcal{G}$. Since (\ref{eq:GNEexistence}) has a solution,
there must exist a GNE for $\mathcal{G}$.

\noindent (b) The proof is based on similar arguments as those in
(a).

\noindent (c) Suppose that $\vect x^{\star}\triangleq\left(\vect p_{q}^{\star},\vect p_{q}^{\texttt{J}\star}\right)_{q=1}^{Q}$
is a GNE of $\mathcal{G}$ satisfying (\ref{eq:linear_const}) for
all $q=1,\ldots,Q$. Then, for every $q$, we have: $\left(\vect p_{q}^{\star},\vect p_{q}^{\texttt{J}\star}\right)\in\mathcal{P}_{q}^{\texttt{sm}}(\vect p_{-q}^{\texttt{J}\star})$
and 
\[
\wt{r_{q}^{s}}\left(\vect p_{q}^{\star},\vect p_{q}^{\texttt{J}\star}\right)=r_{q}^{s}\left(\vect p_{q}^{\star},\vect p_{q}^{\texttt{J}\star}\right)\geq r_{q}^{s}\left(\vect p_{q},\vect p_{q}^{\texttt{J}}\right)=\wt{r_{q}^{s}}\left(\vect p_{q},\vect p_{q}^{\texttt{J}}\right),
\]
for all $\left(\vect p_{q},\vect p_{q}^{\texttt{J}}\right)\in\mathcal{P}_{q}^{\texttt{sm}}(\vect p_{-q}^{\texttt{J}\star})$.
Therefore, $\left(\vect p_{q}^{\star},\vect p_{q}^{\texttt{J}\star}\right)$
is a solution of (\ref{SCproblem-smooth}), with $\vect p_{-q}^{\texttt{J}}=\vect p_{-q}^{\texttt{J}\star}$.

\noindent (d) Suppose that $\vect x^{\star}\triangleq\left(\vect p_{q}^{\star},\vect p_{q}^{\texttt{J}\star}\right)_{q=1}^{Q}$
is a GNE of $\mathcal{G}^{\texttt{sm}}$. Then, for each $q$, $\left(\vect p_{q}^{\star},\vect p_{q}^{\texttt{J}\star}\right)\in\mathcal{P}_{q}(\vect p_{-q}^{\texttt{J}\star})$
and 
\[
r_{q}^{s}\left(\vect p_{q}^{\star},\vect p_{q}^{\texttt{J}\star}\right)=\wt{r_{q}^{s}}\left(\vect p_{q}^{\star},\vect p_{q}^{\texttt{J}\star}\right)\geq\wt{r_{q}^{s}}\left(\vect p_{q},\vect p_{q}^{\texttt{J}}\right)=r_{q}^{s}\left(\vect p_{q},\vect p_{q}^{\texttt{J}}\right),
\]
for all $\left(\vect p_{q},\vect p_{q}^{\texttt{J}}\right)\in\mathcal{P}_{q}^{\texttt{sm}}(\vect p_{-q}^{\texttt{J}\star})$.
Since $r_{q}^{s}\left(\vect p_{q},\vect p_{q}^{\texttt{J}}\right)=0$
for all $\left(\vect p_{q},\vect p_{q}^{\texttt{J}}\right)\in\mathcal{P}_{q}(\vect p_{-q}^{\texttt{J}\star})\setminus\mathcal{P}_{q}^{\texttt{sm}}(\vect p_{-q}^{\texttt{J}\star})$
, we have 
\[
r_{q}^{s}\left(\vect p_{q}^{\star},\vect p_{q}^{\texttt{J}\star}\right)\geq r_{q}^{s}\left(\vect p_{q},\vect p_{q}^{\texttt{J}}\right),\quad\forall\left(\vect p_{q},\vect p_{q}^{\texttt{J}}\right)\in\mathcal{P}_{q}(\vect p_{-q}^{\texttt{J}\star}).
\]
Therefore, $\left(\vect p_{q}^{\star},\vect p_{q}^{\texttt{J}\star}\right)$
is a solution of (\ref{eq:SCproblem}), with $\vect p_{-q}^{\texttt{J}}=\vect p_{-q}^{\texttt{J}\star}$.
\vspace{-0.3cm}

\section{Proof of Proposition \ref{prop:connections_QGNE}\label{sec:Proof-of-Proposition_QGNE}}

\noindent (a) It follows from Proposition~\ref{prop:connections-GNE}(a)
that a GNE of $\mathcal{G}$ always exists; since the players' optimization
problems in $\mathcal{G}$ have polyhedral sets, every GNE of $\mathcal{G}$
is also a B-QGNE. Therefore a B-QGNE of $\mathcal{G}$ exists.

\noindent (b) It follows from Proposition~\ref{prop:connections-GNE}(b)
and similar arguments as in the proof of (a).

\noindent (c) This proof is based on the following fact, regarding
the directional derivative of the plus function \textcolor{black}{\cite[Eq. 2.124]{plusFunctionDerivative}:}
\begin{equation}
\begin{aligned}r_{q}^{s}{'}(\vect z;\vect y-\vect z) & =[\max(0,\wt{r_{q}^{s}}(\bullet))]{'}(\vect z;\vect y-\vect z)\\
 & =\left\{ \setlength\arraycolsep{0.03in}\begin{array}{lc}
\max\left(0,\nabla_{\vect x_{q}}\wt{r_{q}^{s}}(\vect z)^{T}(\vect y-\vect z)\right), & \text{if }\wt{r_{q}^{s}}(\vect z)=0\vspace{0.05in}\\
\nabla_{\vect x_{q}}\wt{r_{q}^{s}}(\vect z)^{T}(\vect y-\vect z), & \text{if }\wt{r_{q}^{s}}(\vect z)>0.
\end{array}\right.
\end{aligned}
\label{eq:direc_derivative}
\end{equation}
Let $\vect x^{\star}\triangleq\left(\vect x_{q}^{\star}\right)_{q=1}^{Q}$
be a (restricted) B-QGNE of $\mathcal{G}$ satisfying (\ref{eq:linear_const})
for all $q=1,\ldots,Q$, with $\vect x_{q}^{\star}\triangleq\left(\vect p_{q}^{\star},\vect p_{q}^{\texttt{J}\star}\right)$.
Then, for every $q$, based on (\ref{eq:direc_derivative}), consider
the following two cases:

\noindent \emph{Case I}:\emph{ $\wt{r_{q}^{s}}(\vect x_{q}^{\star})>0$}.
For all $\vect x_{q}\in\mathcal{P}_{q}^{\texttt{sm}}(\vect p_{-q}^{\texttt{J}\star})$
we have 
\[
0\geq r_{q}^{s}{'}(\mathbf{x}_{q}^{\star};\mathbf{x}_{q}-\mathbf{x}_{q}^{\star})=\nabla_{\vect x_{q}}\wt{r_{q}^{s}}(\vect x_{q}^{\star})^{T}\left(\vect x_{q}-\vect x_{q}^{\star}\right).
\]
\emph{Case II}:\emph{ $\wt{r_{q}^{s}}(\vect x_{q}^{\star})=0$}. For
all $\vect x_{q}\in\mathcal{P}_{q}^{\texttt{sm}}(\vect p_{-q}^{\texttt{J}\star})$
we have 
\[
0\geq r_{q}^{s}{'}(\mathbf{x}_{q}^{\star};\mathbf{x}_{q}-\mathbf{x}_{q}^{\star})\geq\nabla_{\vect x_{q}}\wt{r_{q}^{s}}(\vect x_{q}^{\star})^{T}\left(\vect x_{q}-\vect x_{q}^{\star}\right).
\]
The desired result follows readily from the above two cases.

\noindent (d) Let $\vect x^{\star}\triangleq\left(\vect x_{q}^{\star}\right)_{q=1}^{Q}$
be a QGNE of $\mathcal{G}^{\texttt{sm}}$, with $\vect x_{q}^{\star}\triangleq\left(\vect p_{q}^{\star},\vect p_{q}^{\texttt{J}\star}\right)$.
Consider an arbitrary $q$. Then $\vect x_{q}^{\star}\in\mathcal{P}_{q}(\vect p_{-q}^{\texttt{J}\star})$.
Based on (\ref{eq:direc_derivative}), consider the following two
cases:

\noindent \emph{Case I}:\emph{ $\wt{r_{q}^{s}}(\vect x_{q}^{\star})>0$}.
For all $\vect x_{q}\in\mathcal{P}_{q}^{\texttt{sm}}(\vect p_{-q}^{\texttt{J}\star})$
we have 
\[
r_{q}^{s}{'}(\mathbf{x}_{q}^{\star};\mathbf{x}_{q}-\mathbf{x}_{q}^{\star})=\nabla_{\vect x_{q}}\wt{r_{q}^{s}}(\vect x_{q}^{\star})^{T}\left(\vect x_{q}-\vect x_{q}^{\star}\right)\leq0
\]
\emph{Case II}:\emph{ $\wt{r_{q}^{s}}(\vect x_{q}^{\star})=0$}. For
all $\vect x_{q}\in\mathcal{P}_{q}^{\texttt{sm}}(\vect p_{-q}^{\texttt{J}\star})$
we have 
\[
r_{q}^{s}{'}(\mathbf{x}_{q}^{\star};\mathbf{x}_{q}-\mathbf{x}_{q}^{\star})=\max\left(0,\nabla_{\vect x_{q}}\wt{r_{q}^{s}}(\vect x_{q}^{\star})^{T}\left(\vect x_{q}-\vect x_{q}^{\star}\right)\right)=0
\]
The desired result comes readily from the above two cases.

\vspace{-0.2cm}

\textcolor{black}{{} \bibliographystyle{IEEEtran}
\bibliography{biblioFinal_mod,IEEEabrv}
 } 
\end{document}